\documentclass[journal]{IEEEtran}
\usepackage[noadjust]{cite}

\ifCLASSINFOpdf
\else
\fi
\usepackage{algorithmic}
\usepackage{float}
\floatstyle{ruled}
\newfloat{algorithm}{tbp}{loa}
\providecommand{\algorithmname}{Algorithm}
\floatname{algorithm}{\protect\algorithmname}
\usepackage{pdfsync}
\usepackage{graphicx}
\usepackage{amsmath}
\usepackage{amssymb}
\usepackage{amsfonts}\allowdisplaybreaks
\usepackage{pgfplots}
\usepackage[english]{babel}
\usepackage{subcaption}
\usepackage{blindtext}
\usepackage[normalem]{ulem}
\usepackage{cancel}

\renewcommand{\vec}[1]{\boldsymbol{#1}}
\newcommand{\EV}[1]{\mathbb{E}\left\lbrace #1\right\rbrace}
\newcommand{\realset}{\mathbb{R}}
\newcommand{\complexset}{\mathbb{C}}
\newcommand{\norm}[1]{\left|\left|#1\right|\right|}
\newcommand{\normsca}[1]{\left|#1\right|}
\newcommand{\tr}[1]{\text{tr}\left\lbrace#1\right\rbrace}
\newcommand{\proj}[2]{\vec{P}_{#1}#2}
\newcommand{\oproj}[2]{\vec{P}_{#1}^{\perp}#2}

\newcommand{\iter}[2]{#1^{(#2)}}
\newcommand{\sigs}[1]{\sigma_{\text{s, #1}}}
\newtheorem{theorem}{Theorem}

\newtheorem{Proposition}{Proposition}

\pgfplotsset{compat=1.6}

\begin{document}
%
\title{Partial Relaxation Approach: An Eigenvalue-Based DOA Estimator Framework}
%
%
%

\author{Minh~Trinh-Hoang,
        Mats~Viberg
        and~Marius~Pesavento
\thanks{Part of this work has been accepted for publication in \textit{2017 IEEE International Workshop on Computational Advances in Multi-Sensor Adaptive Processing (CAMSAP 2017)} and \textit{2018 IEEE International Conference on Acoustics, Speech and Signal Processing (ICASSP 2018)}}        
\thanks{Minh Trinh-Hoang and Marius Pesavento are with the Communication
	Systems Group, TU Darmstadt, Darmstadt, Germany
	(e-mail: {thminh, pesavento}@nt.tu-darmstadt.de).}
\thanks{Mats Viberg is with Department of Electrical Engineering, Chalmers University of Technology, Gothenburg, Sweden (e-mail: mats.viberg@chalmers.se).}
\thanks{Manuscript received \today; revised \today.}}

\maketitle
\begin{abstract}
	In this paper, the partial relaxation approach is introduced and applied to the DOA estimation problem using spectral search. Unlike existing spectral-based methods like conventional beamformer, Capon beamformer or MUSIC which can be considered as single source approximation of multi-source estimation criteria, the proposed approach accounts for the existence of multiple sources. At each considered direction, the manifold structure of the remaining interfering signals impinging on the sensor array is relaxed, which results in closed form estimates for the ``interference" parameters. Thanks to this relaxation, the conventional multi-source optimization problem reduces to a simple spectral search. Following this principle, we propose estimators based on the Deterministic Maximum Likelihood, Weighted Subspace Fitting and covariance fitting methods. To calculate the null-spectra efficiently, an iterative rooting scheme based on the rational function approximation is applied to the partial relaxation methods. Simulation results show that, irrespectively of any specific structure of the sensor array, the performance of the proposed estimators is superior to the conventional methods, especially in the case of low Signal-to-Noise-Ratio and low number of snapshots, while maintaining a computational cost which is comparable to MUSIC.
\end{abstract}


%
\IEEEpeerreviewmaketitle

\section{Introduction}
%
%
%
%


Direction-of-Arrival (DOA) estimation and source localization have been fundamental and long-established research directions in sensor array processing. The application of DOA estimation spans multiple fields of research, including wireless communication, radio astronomy, automotive radar, etc. \cite{rembovsky2009radio, van2004detection, twodecades, bookChapter14}.

Many methods for DOA estimation have been developed to increase the resolution capability, computational efficiency and robustness of the algorithms. Although the family of Maximum Likelihood (ML) estimators enjoys remarkable properties of excellent threshold and asymptotic performance \cite{MLPerformance, MLPerformance2, MLandCRB}, the application of ML estimators in real-time scenarios is generally impractical due to the optimization of multi-modal functions and the associated prohibitive computational cost. In the family of subspace-based algorithms, MUSIC \cite{music} relies on the signal subspace calculated from the spatial sample covariance matrix and performs a spectral search for the estimated DOAs. In \cite{RMTMusic}, a modified version of the MUSIC algorithm based on the Random Matrix Theory is proposed to improve the threshold performance. On the other hand, root-MUSIC \cite{rootmusic}, ESPRIT \cite{esprit} and their unitary variants \cite{unitaryROOTMUSIC}, \cite{EPUMA}, \cite{unitaryESPRPIT} exploit uniform linear and shift-invariant array structures, respectively, to provide search-free DOA estimates, resulting in considerable reduction in the computational time and enhancement in the estimation performance \cite{rootMUSICPerformance,rootMUSICandMINNORM,ESPRITPerformance,bookChapter15}. 

When formulated as non-linear least squares (LS) problems, conventional spectral-based algorithms ignore the existence of multiple sources in the snapshots and therefore can be regarded as single source approximation of multi-source criteria \cite{MLPerformance2}, \cite{PAULRAJ1993693}. As a consequence, if the interference power from other sources is high, the performance of conventional algorithms strongly degrades \cite{MLPerformance}, \cite{MVDRPerformance}. This scenario occurs, e.g., when two or multiple sources are closely-spaced.

To overcome the aforementioned shortcomings of existing estimators without requiring specific structures of the sensor array, in this paper, DOA estimators based on the partial relaxation approach \cite{Trinh-Hoang2017, Trinh-Hoang2018} are presented. Taking a fundamentally different perspective from the conventional spectral-based algorithms, the partial relaxation approach takes signals from both ``desired" and ``interfering" directions into account. However, while the manifold structure of the desired direction is unaltered, the manifold structure of the interfering directions is relaxed to make the problem computationally tractable, hence the name partial relaxation. Based on this concept, closed-form expressions for the optimal solutions of the relaxed interference parameters are first determined, and then substituted back into the multi-source criteria, resulting in simple spectral search procedures. In contrast to MUSIC, in which the eigenvectors spanning the noise subspace play an essential role in the calculation of the null-spectrum, the partial relaxation approach relies only on the eigenvalues of a certain modified covariance matrix at each direction. In comparison to the corresponding conventional multi-source fitting methods, the partial relaxation approach admits simpler solutions while obtaining superior error performance to the conventional spectral-based algorithms. To summarize, the original contributions of this paper are:
\begin{itemize}
	\item We introduce a new \textit{Partial Relaxation Framework} for the DOA estimation problem, which, from the simulation results, exhibits excellent Signal-to-Noise (SNR) threshold performance without requiring any particular structure of the sensor array.
	\item We propose four new DOA estimators under the partial relaxation framework based on the classical Deterministic Maximum Likelihood, Weighted Subspace Fitting, constrained and unconstrained covariance fitting estimator.
	\item In order to reduce the overall computational complexity, we propose an efficient procedure for computing the required null-spectra of the proposed estimators under the partial relaxation framework.
\end{itemize}

The paper is organized as follows. The signal model is introduced in Section~\ref{sec:SignalMode}. Existing DOA methods based on non-linear least squares problems, which are the motivating background of the proposed work, are introduced in Section~\ref{sec:ConvNLS}. The mathematical formulation of the proposed partial relaxation approach and its adaptation to the conventional DOA estimation methods, i.e., the Deterministic ML, Weighted Subspace Fitting, constrained and unconstrained covariance fitting estimator, are described in Section~\ref{sec:PRMethod}. The computational aspects of the partial relaxation framework are discussed in Section~\ref{sec:computationalAspect}, where the rational approximation is applied to calculate the eigenvalues efficiently and therefore avoid the full computation of the eigenvalue decomposition. To illustrate the performance gain in terms of estimation errors and execution time of the proposed methods, simulation results based on synthetic data are presented in Section~\ref{sec:SimResult}. Lastly in Section~\ref{sec:Conclusion}, remarks and extensions to further research are discussed.

\textbf{Notation:} Matrices are denoted by boldface uppercase letters $\vec{A}$, vectors are denoted by boldface lowercase letters $\vec{a}$, and scalars are denoted by regular letters $a$. $\vec{I}_M$ represents the $M\times M$ identity matrix. Symbols $(\cdot)^H$, $(\cdot)^{-1}$ and $(\cdot)^{1/2}$ denote the Hermitian transpose, inverse and the principal square root, respectively, of the matrix argument. The expectation operator is represented by $\EV{\cdot}$. The trace operator is denoted by $\tr{\cdot}$, and the determinant is represented by $\text{det}(\cdot)$. $\norm{\cdot}_\text{F}$ denotes the Frobenius norm, and $\norm{\cdot}_2$ is the $\ell_2$-norm of the argument. Finally, $^N\arg\min f(\cdot)$ denotes the $N$ arguments at which the function $f(\cdot)$ attains its $N$-deepest separated local minima.
\section{Signal Model}\label{sec:SignalMode}
Consider an array of $M$ sensors receiving $N$ narrowband signals emitted from sources with corresponding unknown DOAs $\vec{\theta} = \left[\theta_1, \ldots, \theta_N\right]^T$. Furthermore, assume that $N < M$. The sensor measurement vector ${\vec{x}(t) = \left[x_1(t), \ldots, x_M(t)\right]^T\in\complexset^{M\times 1}}$ in the baseband at the time instant $t$ is modeled as:
\begin{equation}
\label{signal_model}
\vec{x}(t)= \vec{A(\theta)}\vec{s}(t)+\vec{n}(t) \text{ with } t = 1, \ldots, T,
\end{equation}
where $\vec{s}(t) = \left[ s_1(t), \ldots, s_N(t)\right]^T\in\complexset^{N\times 1}$ denotes the baseband source signal vector from $N$ sources and ${\vec{n}(t)\in\complexset^{M\times 1}}$ represents the additive circularly complex noise vector at the sensor array with the noise covariance matrix ${\EV{\vec{n}(t)\vec{n}(t)^H}=\sigma_{\text{n}}^2\vec{I}_M}$. The steering matrix ${\vec{A}(\vec{\theta})\in\complexset^{M\times N}}$ in \eqref{signal_model}, which is assumed to have full column rank, is given by:
\begin{equation}
\vec{A}(\vec{\theta}) = \left[\vec{a}(\theta_1), \ldots, \vec{a}(\theta_N)\right],
\end{equation}
where $\vec{a}(\theta_{\text{n}})$ denotes the sensor array response for the DOA $\theta_{\text{n}}$. Equation~\eqref{signal_model} can be rewritten for multiple snapshots ${t=1, \ldots, T}$ in a compact notation as:
\begin{equation}
\label{eq:multisamples}
\vec{X} = \vec{A}(\vec{\theta})\vec{S} + \vec{N},
\end{equation}
where $\vec{X}=\left[\vec{x}(1), \ldots, \vec{x}(T)\right]\in\complexset^{M\times T}$ is the received baseband signal matrix. In a similar manner, we define the source signal matrix $\vec{S}\in\complexset^{N\times T}$ and the sensor noise matrix ${\vec{N}\in\complexset^{M\times T}}$ as ${\vec{S}=\left[\vec{s}(1), \ldots, \vec{s}(T)\right]}$ and ${\vec{N}=\left[\vec{n}(1), \ldots, \vec{n}(T)\right]}$, respectively. 

Assume that the source signals and the noise are uncorrelated, the covariance matrix of the received signal $\vec{R}\in\complexset^{M\times M}$ is given by: 
\begin{equation}
\vec{R} = \EV{\vec{x}(t)\vec{x}(t)^H} = \vec{A}\vec{R}_{\text{s}}\vec{A}^H + \sigma_{\text{n}}^2\vec{I}_M,
\label{eq:covarianceMatrix}
\end{equation}
where $\vec{R}_{\text{s}} =\EV{\vec{s}(t)\vec{s}(t)^H}$ is the covariance matrix of the transmitted signal $\vec{s}(t)$. We assume throughout the paper that the number of sources $N$ is known, and the source signals are non-coherent.

In practice, the true covariance matrix $\vec{R}$ is not available and the sample covariance matrix $\hat{\vec{R}}$ is used instead:
\begin{equation}
\label{eq:defCovMat}
\hat{\vec{R}} = \dfrac{1}{T}\vec{X}\vec{X}^H.
\end{equation}
Subspace techniques rely on the properties of the eigenspaces of the sample covariance matrix $\hat{\vec{R}}$, which is decomposed as:
\begin{subequations}
\begin{align}
\label{eq:EVD}\hat{\vec{R}} &= \hat{\vec{U}}\hat{\vec{\Lambda}}\hat{\vec{U}}^H\\
\label{eq:eigdecsample}&=\hat{\vec{U}}_\text{s}\hat{\vec{\Lambda}}_{\text{s}}\hat{\vec{U}}_{\text{s}}^H + \hat{\vec{U}}_{\text{n}}\hat{\vec{\Lambda}}_{\text{n}}\hat{\vec{U}}_{\text{n}}^H.
\end{align}
\end{subequations}
In \eqref{eq:eigdecsample}, $\hat{\vec{\Lambda}}_{\text{s}} \in\complexset^{N\times N}$ is a diagonal matrix, containing the $N$-largest eigenvalues $\lbrace\hat{\lambda}_1, \ldots, \hat{\lambda}_N\rbrace$, and $\hat{\vec{U}}_{\text{s}}\in\complexset^{M\times N}$ contains the corresponding $N$-principal eigenvectors of the sample covariance matrix $\hat{\vec{R}}$. Similarly, ${\hat{\vec{\Lambda}}_{\text{n}}\in \complexset^{(M-N)\times(M-N)}}$ and ${\hat{\vec{U}}_{\text{n}}\in\complexset^{M\times(M-N)}}$ contain the $(M-N)$-noise eigenvalues $\lbrace\hat{\lambda}_{N+1},\ldots, \hat{\lambda}_M\rbrace$ and the associated noise eigenvectors, respectively.
\section{Existing Methods based on Non-linear LS}\label{sec:ConvNLS}
In the family of ML estimators, the Deterministic ML (DML) estimates the DOAs by searching for the steering matrix $\vec{A}$ in the $N$-source array manifold $\mathcal{A}_N$, which is parameterized as follows:
\begin{equation}
\mathcal{A}_N = \left\lbrace\vec{A}| \vec{A} = \left[\vec{a}(\vartheta_1), \ldots, \vec{a}(\vartheta_N)\right],\vartheta_1<\ldots<\vartheta_N\right\rbrace.
\label{eq: paramterizedArrayManifold}
\end{equation}
Based on the signal model in \eqref{eq:multisamples} and the parameterization in \eqref{eq: paramterizedArrayManifold}, the DML estimator is formulated as the following non-linear least squares problem \cite{van2004detection}:
\begin{equation}
\left\lbrace\hat{\vec{A}}_{\text{DML}}, \hat{\vec{S}} \right\rbrace = \underset{\vec{A}\in\mathcal{A}_N, \vec{S}\in\complexset^{N\times T}}{\arg\min}\text{ }\norm{\vec{X}- \vec{A}\vec{S}}_\text{F}^2.
\label{eq: DML1}
\end{equation}
In the case that only the DOAs are considered, the DML estimator in \eqref{eq: DML1} can be reformulated as:
\begin{equation}
\left\lbrace\hat{\vec{A}}_{\text{DML}}\right\rbrace = \underset{\vec{A}\in\mathcal{A}_N}{\arg\min} \text{ } \tr{\oproj{\vec{A}}{\hat{\vec{R}}}}.
\label{eq: DML2}
\end{equation}
In \eqref{eq: DML2}, $\proj{\vec{A}}{} = \vec{A}\left(\vec{A}^H\vec{A}\right)^{-1}\vec{A}^H$ denotes the projection matrix onto the subspace spanned by the columns of the matrix $\vec{A}$. Similarly, $\oproj{\vec{A}}{} = \vec{I}_M - \proj{\vec{A}}{}$ is the projection matrix onto the subspace which is the orthogonal complement to the subspace $\text{span}(\vec{A})$.

In general, the DOA estimation problem can be formulated as:
\begin{equation}
\label{eq:generalDOAProblem}
\left\lbrace\hat{\vec{A}}\right\rbrace = \underset{\vec{A}\in\mathcal{A}_N}{\arg\min } \text{ }f\left(\vec{A}, \vec{Y}\right),
\end{equation}
where $f(\cdot)$ denotes a general cost function, and $\vec{Y}$ is a data matrix which is somehow obtained from the received baseband signal matrix $\vec{X}$. We remark that different choices on the cost function $f(\cdot)$, the parameterization of $\vec{A}$ and the data matrix $\vec{Y}$ result in different error performance of the DOA estimators. Since the $N$-source array manifold $\mathcal{A}_N$ is highly structured and non-convex, the optimization problem in \eqref{eq:generalDOAProblem} is generally challenging \cite{MV_WSF,Ottersten1993,AlternatingOptimization}. To relieve the high computational cost, a common approach is to find a sub-optimal solution of \eqref{eq:generalDOAProblem} by considering a special case: the single source approximation \cite{PAULRAJ1993693}, \cite{stoica2005spectral}. In this approach, we consider only the single source array manifold as the feasible set of the optimization problem in \eqref{eq:generalDOAProblem}, i.e., $\vec{A}=\vec{a}\in\mathcal{A}_1$, while leaving the data matrix $\vec{Y}$ unchanged. The locations of $N$-deepest minima of the null-spectrum $f\left(\vec{a}, \vec{Y}\right)$, which are obtained by performing a sweep search on $\vec{a}\in\mathcal{A}_1$, correspond to the steering vectors of the estimated DOAs.	The above mentioned steps are compactly expressed by the following notation:
\begin{equation}
\left\lbrace\hat{\vec{a}}\right\rbrace = \text{ }^N \underset{\vec{a}\in\mathcal{A}_1}{\arg\min } \text{ }f\left(\vec{a}, \vec{Y}\right).
\end{equation}
From now on, unless we want to emphasize the dependence of the steering vector $\vec{a}(\vartheta)$ on the direction, the argument $\vartheta$ will be omitted. Under the single source approximation approach, conventional spectral-search DOA estimators from the literature are retrieved by considering different optimizing criteria and different data matrices \cite{PAULRAJ1993693}:
\begin{itemize}
	\item \textbf{Measurement Fitting}:
	Using the cost function of the DML in~\eqref{eq: DML2} and the data matrix $\vec{Y} = \hat{\vec{R}}$ under the single source approximation, the following optimization problem is obtained:
	\begin{equation}
	\left\lbrace\hat{\vec{a}}\right\rbrace =\text{ }^N \underset{\vec{a}\in\mathcal{A}_1}{\arg\min}\text{ }  \tr{\oproj{\vec{a}}{\hat{\vec{R}}}}.
	\label{eq:BF}
	\end{equation}

	Note that the objective function in \eqref{eq:BF} is the null-spectrum of the conventional beamformer \cite{van2004detection}.
	\item \textbf{Weighted Subspace Fitting (WSF)}:
	In accordance with the DML method, the optimization problem for the Weighted Signal Subspace Fitting is formulated in \cite{MV_WSF} as:
	\begin{equation}
	\left\lbrace\hat{\vec{A}}\right\rbrace = \underset{\vec{A}\in\mathcal{A}_N}{\arg\min}\text{ }\tr{\oproj{\vec{A}}{\hat{\vec{U}}_{\text{s}}\vec{W}\hat{\vec{U}}}_{\text{s}}^H},\label{eq:FullWSF}
	\end{equation}
	where $\vec{W}\in\complexset^{N\times N}$ is a positive semidefinite weighting matrix. In \cite{MV_WSF}, the authors showed that by choosing the weighting matrix as:
	\begin{equation}
	\label{eq:optimalWeighting_WSF}
	\vec{W} = \hat{\tilde{\vec{\Lambda}}}^2\hat{\vec{\Lambda}}_{\text{s}}^{-1}\vspace*{-6pt}
	\end{equation} with $\hat{\tilde{\vec{\Lambda}}} = \hat{\vec{\Lambda}}_{\text{s}} - \hat{\sigma}_{\text{n}}^2\vec{I}_N$ and $\hat{\sigma}_{\text{n}}^2 = \dfrac{1}{M - N}\sum\limits_{k = N+1}^{M}\hat{\lambda}_k$, the estimation error of the WSF method asymptotically achieves the Cramer-Rao Bound as the number of snapshots $T$ tends to infinity. When the single source approximation is adopted with the data matrix $\vec{Y} = \hat{\vec{U}}_{\text{s}}\vec{W}\hat{\vec{U}}_{\text{s}}^H$, we obtain the following optimization problem:
	\begin{equation}
	\left\lbrace\hat{\vec{a}}\right\rbrace =\text{ }^N\underset{\vec{a}\in\mathcal{A}_1}{\arg\min}  \text{ }\tr{\oproj{\vec{a}}\hat{\vec{U}}_{\text{s}}\vec{W}\hat{\vec{U}}_{\text{s}}^H}.
	\label{eq:WSF}
	\end{equation}

	In a special case when $\vec{W} = \vec{I}_N$, the formulation in \eqref{eq:WSF} can be shown to be equivalent to the MUSIC estimator \cite{PAULRAJ1993693}. 
	\item \textbf{Covariance Fitting}: Starting from the identity in \eqref{eq:covarianceMatrix} and applying the least squares covariance fitting without considering the weighting matrix $\vec{W}$, we obtain \cite{COMET}:
	\begin{equation}
	\label{eq:fullCF}
	\left\lbrace\hat{\vec{A}}, \hat{\vec{R}}_{\text{s}}\right\rbrace = \underset{\vec{A}\in\mathcal{A}_N, \vec{R}_{\text{s}} \succeq \vec{0}}{\arg\min}\norm{\hat{\vec{R}} - \vec{A}\vec{R}_{\text{s}}\vec{A}^H}_\text{F}^2.
	\end{equation}

	If the single source consideration is adopted on \eqref{eq:fullCF}  with the data matrix $\vec{Y} = \hat{\vec{R}}$, we obtain:
\begin{equation}
	\left\lbrace\hat{\vec{a}}\right\rbrace =\text{ }^N\underset{\vec{a}\in\mathcal{A}_1}{\arg\min}  \text{ }\underset{\sigma_{\text{s}}^2\geq 0}{\min}\norm{\hat{\vec{R}}-\sigma_{\text{s}}^2\vec{a}\vec{a}^H}_\text{F}^2.
	\label{eq:CMFull}
	\end{equation}
	The inner optimization problem in \eqref{eq:CMFull} obtains a closed-form minimizer $\hat{\sigma}_{\text{s}}^2$ given by \cite{stoica2005spectral}:
	\begin{equation}
	\hat{\sigma}_{\text{s}}^2 = \underset{\sigma_{\text{s}}^2\geq 0}{\arg\min}\norm{\hat{\vec{R}}-\sigma_{\text{s}}^2\vec{a}\vec{a}^H}_\text{F}^2 = \dfrac{\vec{a}^H\hat{\vec{R}}\vec{a}}{(\vec{a}^H\vec{a})^2},
	\label{eq:CM1}
	\end{equation}
	which is merely a scaled spectrum of the conventional beamformer. If a positive semidefinite constraint is enforced in the inner optimization problem in \eqref{eq:CM1} as:
	\begin{equation}
	\begin{aligned}
	\hat{\sigma}_{\text{s}}^2 = \underset{\sigma_{\text{s}}^2\geq 0}{\arg\min} &\norm{\hat{\vec{R}} - \sigma_{\text{s}}^2\vec{a}\vec{a}^H}_\text{F}^2\\
	\text{subject to } &\hat{\vec{R}} - \sigma_{\text{s}}^2\vec{a}\vec{a}^H \succeq \vec{0},
	\label{eq:CaponFormulation}
	\end{aligned}
	\end{equation}
	then the minimizer $\hat{\sigma}_{\text{s}}^2$ is given by the Capon spectrum \cite[p. 293]{stoica2005spectral}, \cite{RobustCaponBeamformer}: 
	\begin{equation}
	\label{eq:CaponSpectrum}
	\hat{\sigma}_{\text{s}}^2 = \dfrac{1}{\vec{a}^H\hat{\vec{R}}^{-1}\vec{a}}.
	\end{equation}
\end{itemize}

We remark that the optimization problems obtained by applying the single source approximation approach to the least squares fitting problems in \eqref{eq: DML2}, \eqref{eq:FullWSF} and \eqref{eq:fullCF} are equivalent to the multi-source criteria counterparts under the assumptions of orthogonal steering vectors, i.e., ${\vec{A}^H\vec{A}=M\vec{I}_N}$, and uncorrelated source signals.
In this case, the effects of interfering signals on the desired direction vanish. As a result, the steering vectors are decoupled, and the multi-source optimization problems can be decomposed into multiple single source estimation problems. Conversely, if the steering vectors are not orthogonal, the performance of the single source approach degrades due to the interference of neighboring source signals.
\section{Partial Relaxation Approach}\label{sec:PRMethod}
In order to relieve the aforementioned drawbacks of the conventional spectral-search algorithms, in this section, the general concept for the partial relaxation approach is introduced. Afterwards, four DOA estimators are proposed by adopting the partial relaxation approach on the classical least squares problems in \eqref{eq: DML2}, \eqref{eq:FullWSF}, \eqref{eq:fullCF} and \eqref{eq:CaponFormulation}.
\subsection{General Concept}\label{sec:PRConcept}
Unlike the single source approximation, our proposed partial relaxation approach considers the signals from both the ``desired" and ``interfering" directions. However, to make the problem tractable, the array structures of the interfering signals are relaxed. More precisely, instead of enforcing the steering matrix ${\vec{A} = \left[\vec{a}(\theta_1), \ldots, \vec{a}(\theta_N)\right]}$ to be an element in the highly structured array manifold $\mathcal{A}_N$ as in \eqref{eq:generalDOAProblem}, without the loss of generality, we maintain the manifold structure of the first column $\vec{a}(\theta_1)$ of $\vec{A}$, which corresponds to the signal of consideration. On the other hand, the manifold structure of the remaining sources $\left[\vec{a}(\theta_2), \ldots, \vec{a}(\theta_N)\right]$, which are considered as interfering sources, is relaxed to an arbitrary matrix ${\vec{B}\in\complexset^{M\times(N-1)}}$. Mathematically, we assume that $\vec{A}\in\bar{\mathcal{A}}_N$, where the relaxed array manifold $\bar{\mathcal{A}}_N$ is parameterized as:
\begin{equation}
\bar{\mathcal{A}}_N = \left\lbrace\vec{A}| \vec{A} = \left[\vec{a}(\vartheta), \vec{B}\right], \vec{a}(\vartheta)\in\mathcal{A}_1, \vec{B}\in\complexset^{M\times(N-1)}\right\rbrace.
\label{eq: relaxedArrayManifold}
\end{equation}
Note that $\bar{\mathcal{A}}_N$ still retains some structure depending on the geometry of the sensor array, hence the name partial relaxation. However, only one DOA can be estimated if the cost function of \eqref{eq:generalDOAProblem} is minimized on the relaxed array manifold $\bar{\mathcal{A}}_N$ of \eqref{eq: relaxedArrayManifold}. Therefore, the grid search is applied similarly to the single source approximation in Section~\ref{sec:ConvNLS} as follows: first we fix the data matrix $\vec{Y}$, minimize the objective function in \eqref{eq:generalDOAProblem} with respect to $\vec{B}$, and then perform a grid search on $\vec{a}(\vartheta)\in\mathcal{A}_1$ to determine the locations of $N$-deepest local minima. 
The rationale for the partial relaxation approach is that, each time a candidate DOA $\vartheta$ coincides with one of the true DOAs $\theta_n$, then with $\vec{B}$ modeling all other steering vectors, a perfect fit to the data is attained at high SNR or large $T$. When $\vartheta$ is different from all true DOAs, the number of degrees-of-freedom in $\vec{B}$ is not sufficiently large to match to the data perfectly.
In the following, the partial relaxation approach is applied to the four algorithms introduced in Section~\ref{sec:ConvNLS}, i.e., the DML, WSF, constrained and unconstrained covariance fitting estimator. 
\subsection{Partially-Relaxed DML (PR-DML)}\label{subsec:derivationPRDML}
Adopting the partial relaxation approach on the objective function in \eqref{eq: DML2} leads to the following optimization problem:
\begin{equation}
\left\lbrace\hat{\vec{a}}_{\text{PR-DML}}\right\rbrace = \text{ }^N\underset{\vec{a}\in\mathcal{A}_1}{\arg\min}\text{ }\underset{\vec{B}}{\min} \text{ } \tr{\oproj{\left[\vec{a}, \vec{B}\right]}{\hat{\vec{R}}}}.
\label{eq: SDML1}
\end{equation}
By rewriting the objective function in \eqref{eq: SDML1} to decouple $\vec{a}$ and $\vec{B}$ partially, we obtain: 
\begin{equation}
\hspace*{-3pt}
\tr{\oproj{\left[\vec{a}, \vec{B}\right]}{\hat{\vec{R}}}} = \tr{\oproj{\vec{a}}{\hat{\vec{R}}}} - \tr{\proj{\oproj{\vec{a}}{\vec{B}}}{\hat{\vec{R}}}},\label{eq:lasteq}
\end{equation}
where we use the convention that $\tr{\proj{\oproj{\vec{a}}{\vec{B}}}{\hat{\vec{R}}}} = 0$ if $\oproj{\vec{a}}{\vec{B}} = \vec{0}$. Since the first term on the right hand side of \eqref{eq:lasteq} does not depend on $\vec{B}$, the inner optimization problem in \eqref{eq: SDML1} is equivalent to:
\begin{equation}
\underset{\vec{B}}{\max }\text{ } \tr{\proj{\oproj{\vec{a}}{\vec{B}}}{\hat{\vec{R}}}}.\label{eq:immediateStep}
\end{equation}
The solution of \eqref{eq:immediateStep} is given by (see Appendix~\ref{appsec:PRDMLproof}):
\begin{equation}
\label{eq:mainPRDMLresult}
	\underset{\vec{B}}{\max }\text{ } \tr{\proj{\oproj{\vec{a}}{\vec{B}}}{\hat{\vec{R}}}} = \sum\limits_{k = 1}^{N-1}\lambda_k\left(\oproj{\vec{a}}{\hat{\vec{R}}}\right),
\end{equation}
where $\lambda_k(\cdot)$ denotes the $k$-th largest eigenvalue of the matrix in the argument. Substituting \eqref{eq:lasteq} and \eqref{eq:mainPRDMLresult} back into the objective function of \eqref{eq: SDML1}, the null-spectrum of the PR-DML estimator is obtained as:
\begin{equation} \label{eq:PR-DMLPseudospectrum}
\begin{aligned}
f_{\text{PR-DML}}(\vartheta) &= \underset{\vec{B}}{\min}\text{ }\tr{\oproj{\left[\vec{a}(\vartheta), \vec{B}\right]}{\hat{\vec{R}}}}\\
&= \sum\limits_{k=N}^{M}\lambda_k(\oproj{\vec{a}(\vartheta)}{\hat{\vec{R}}}).
\end{aligned}
\end{equation}
The estimated DOAs $\hat{\vec{\theta}} = \left[\hat{\theta}_1, \ldots, \hat{\theta}_N\right]^T$ are then determined by choosing the locations of the $N$-deepest minima of the null-spectrum $f_\text{PR-DML}(\vartheta)$. As further elaborated in Section~\ref{sec:computationalAspect}, the null-spectrum $f_\text{PR-DML}(\vartheta)$ in \eqref{eq:PR-DMLPseudospectrum} can be efficiently computed without necessarily performing a full eigenvalue decomposition at each direction, i.e., computing the corresponding set of eigenvectors of the matrix $\oproj{\vec{a}(\vartheta)}{\hat{\vec{R}}}$ is not required.
\subsection{Partially-Relaxed WSF (PR-WSF)}\label{subsec:PRWSF}
Following a similar derivation as for the PR-DML estimator in Section~\ref{subsec:derivationPRDML} and using the mathematical formulation of the WSF estimator in \eqref{eq:WSF}, the optimization problem corresponding to the PR-WSF estimator reads:
\begin{equation}
\left\lbrace\hat{\vec{a}}_{\text{PR-WSF}}\right\rbrace = \text{ }^N\underset{\vec{a}\in\mathcal{A}_1}{\arg\min}\text{ }\underset{\vec{B}}{\min} \text{ }   \tr{\oproj{\left[\vec{a}, \vec{B}\right]}{\hat{\vec{U}}}_{\text{s}}\vec{W}\hat{\vec{U}}_{\text{s}}^H},
\label{eq:SWSF}
\end{equation}
and the null-spectrum of the PR-WSF is calculated as:
\begin{equation}
\label{eq:PR-WSFPseudospectrum}
f_{\text{PR-WSF}}(\vartheta) = \sum\limits_{k=N}^{M}\lambda_k(\oproj{\vec{a}(\vartheta)}{\hat{\vec{U}}}_{\text{s}}\vec{W}\hat{\vec{U}}_{\text{s}}^H).
\end{equation}
Note that in a special case when $\vec{W} = \vec{I}_N$, the proposed estimator in \eqref{eq:SWSF} is equivalent to the MUSIC estimator (see Appendix~\ref{appsec:PRSF}). From now on, if not further specified, the weighting matrix $\vec{W}$ is chosen as in \eqref{eq:optimalWeighting_WSF}.
\subsection{Partially-Relaxed Constrained Covariance Fitting (PR-CCF)}\label{subsectionPRCCF}
To derive new estimators based on the covariance fitting problems in \eqref{eq:CM1} and \eqref{eq:CaponFormulation}, we follow the principle of the partial relaxation approach for the array steering matrix by relaxing $\vec{A} = \left[\vec{a}, \vec{B}\right]$ with an arbitrary matrix $\vec{B}\in\complexset^{M\times\left(N-1\right)}$. Similarly, we partition the waveform matrix $\vec{S} = \left[\vec{s}, \vec{J}^T\right]^T$ in the signal model in \eqref{eq:multisamples} with $\vec{s}\in\complexset^{T\times 1}$ and $\vec{J} \in\complexset^{(N-1)\times T}$ to obtain:
\begin{equation}
\vec{X}= \vec{a}\vec{s}^T + \vec{E} + \vec{N},
\end{equation}
where $\vec{E}=\vec{B}\vec{J}\in\complexset^{M\times T}$ models the received signal of the remaining $(N-1)$-sources with the relaxed array manifold structure and therefore $\text{rank}(\vec{E}) \leq N-1$. Furthermore, we assume that the sample covariance matrix $\hat{\vec{R}}$ is positive definite, and the signals from other sources are uncorrelated with the signals from the direction $\vec{a}$. Similar to the Capon beamformer in \eqref{eq:CaponFormulation}, the partially-relaxed constrained covariance fitting (PR-CCF) problem is formulated as follows:
\begin{gather}\label{eq:mainCPBased}
\hspace*{-11pt}
\begin{aligned}
\left\lbrace\hat{\vec{a}}_{\text{PR-CCF}}\right\rbrace = \text{ }^N\underset{\vec{a}\in\mathcal{A}_1}{\arg\min}\underset{\sigma_{\text{s}}^2\geq 0,\vec{E}}{\min} &\norm{\hat{\vec{R}} - \sigma_{\text{s}}^2\vec{a}\vec{a}^H - \vec{E}\vec{E}^H}_\text{F}^2\\
\text{subject to } &\hat{\vec{R}} - \sigma_{\text{s}}^2\vec{a}\vec{a}^H - \vec{E}\vec{E}^H \succeq \vec{0}\\
\text{\color{white}subject to } &\text{rank}(\vec{E})\leq N-1.
\end{aligned}\raisetag{0.75\baselineskip}
\end{gather}
Keeping $\left\lbrace\sigma_{\text{s}}^2, \vec{a}\right\rbrace$ fixed and minimizing the objective function of \eqref{eq:mainCPBased} with respect to $\vec{E}$, a low-rank approximation problem is obtained as \cite{HAYDEN1988115}:\vspace*{-8pt}
\begin{gather}
\label{eq:InterferenceOptimization}
\hspace*{-0pt}\begin{aligned}
&\underset{\vec{E}}{\min}\left|\left|\hat{\vec{R}}-\sigma_{\text{s}}^2\vec{a}\vec{a}^H-\vec{E}\vec{E}^H\right|\right|_\text{F}^2 = \sum\limits_{k = N}^{M} \lambda_k^2(\hat{\vec{R}}-\sigma_{\text{s}}^2\vec{a}\vec{a}^H).\\
&\text{subject to } \text{rank}(\vec{E})\leq N-1\raisetag{1\baselineskip}
\end{aligned}
\end{gather}
By performing the eigenvalue decomposition:
\begin{equation}
\hat{\vec{R}} - \sigma_{\text{s}}^2\vec{a}\vec{a}^H = \tilde{\vec{U}}_{\text{s}}\tilde{\vec{\Lambda}}_{\text{s}}\tilde{\vec{U}}_{\text{s}}^H + \tilde{\vec{U}}_{\text{n}}\tilde{\vec{\Lambda}}_{\text{n}}\tilde{\vec{U}}_{\text{n}}^H,
\label{eq:evd_prccf}
\end{equation}
where $\tilde{\vec{\Lambda}}_{\text{s}}$ and $\tilde{\vec{U}}_{\text{s}}$ contain the $(N-1)$ - largest eigenvalues and the corresponding principal eigenvectors of $\hat{\vec{R}}-\sigma_{\text{s}}^{2}\vec{a}\vec{a}^H$, respectively, a minimizer $\hat{\vec{E}}$ of \eqref{eq:InterferenceOptimization} satisfies:
\begin{equation}
\hat{\vec{E}}\hat{\vec{E}}^H = \tilde{\vec{U}}_{\text{s}}\tilde{\vec{\Lambda}}_{\text{s}}\tilde{\vec{U}}_{\text{s}}^H.
\label{eq:minimizerD_prccf}
\end{equation}
Substituting the minimizer $\hat{\vec{E}}$ in \eqref{eq:minimizerD_prccf} back to the inner optimization problem in \eqref{eq:mainCPBased}, we observe that, the constraint $\hat{\vec{R}}-\sigma_{\text{s}}^{2}\vec{a}\vec{a}^H - \hat{\vec{E}}\hat{\vec{E}}^H~\succeq~\vec{0}$ implies that $\hat{\vec{R}}-\sigma_{\text{s}}^{2}\vec{a}\vec{a}^H~\succeq~\vec{0}$. Conversely, for each $\sigma_{\text{s}}^2\geq 0$, if $\hat{\vec{R}}-\sigma_{\text{s}}^{2}\vec{a}\vec{a}^H~\succeq~\vec{0}$, from \eqref{eq:evd_prccf} and \eqref{eq:minimizerD_prccf}, we conclude that $\hat{\vec{R}} - \sigma_{\text{s}}^2\vec{a}\vec{a}^H - \hat{\vec{E}}\hat{\vec{E}}^H~\succeq~\vec{0}$.
As a consequence, an equivalent formulation of the inner problem in \eqref{eq:mainCPBased} is obtained as follows:
\begin{equation}
\label{eq:reformulatedCP}
\begin{aligned}
\underset{\sigma_{\text{s}}^2\geq 0}{\min} &\sum\limits_{k = N}^{M} \lambda_k^2(\hat{\vec{R}} - \sigma_{\text{s}}^2\vec{a}\vec{a}^H)\\
\text{subject to } &\hat{\vec{R}} - \sigma_{\text{s}}^2\vec{a}\vec{a}^H \succeq \vec{0}.
\end{aligned}
\end{equation}
It can be easily shown from the Weyl's inequality regarding the eigenvalues of the modified Hermitian matrix \cite{HORN199829} and the positive semidefiniteness of $\hat{\vec{R}} - \sigma_{\text{s}}^2\vec{a}\vec{a}^H$ that, as long as the constraint in \eqref{eq:reformulatedCP} is not violated, the objective function in \eqref{eq:reformulatedCP} is strictly decreasing as $\sigma_{\text{s}}^2$ increases. Therefore, $\hat{\sigma}_{\text{s, C}}^2$ is a minimizer of \eqref{eq:reformulatedCP} if and only if the matrix $\hat{\vec{R}} - \hat{\sigma}_{\text{s, C }}^2\vec{a}\vec{a}^H$ possesses at least one eigenvalue equal to zero. Consequently, the minimizer $\hat{\sigma}_{\text{s, C }}^2$ is obtained from the Capon spectrum \cite[Equation (6.5.33)]{stoica2005spectral}:
\begin{equation}\label{eq:CPSpec}
\hat{\sigma}_{\text{s, C}}^2 = \dfrac{1}{\vec{a}^H\hat{\vec{R}}^{-1}\vec{a}}.
\end{equation}
Substitute \eqref{eq:reformulatedCP} and \eqref{eq:CPSpec} back into \eqref{eq:mainCPBased}, the PR-CCF estimator returns the estimated DOAs by determining the $N$-deepest minima of the following null-spectrum:
\begin{equation}
\label{eq:PR-CCFPseudospectrum}
f_\text{PR-CCF}(\vartheta) = \sum\limits_{k=N}^{M}\lambda_k^2\mathopen{}\left(\hat{\vec{R}} - \dfrac{1}{\vec{a}(\vartheta)^H\hat{\vec{R}}^{-1}\vec{a}(\vartheta)}\vec{a}(\vartheta)\vec{a}(\vartheta)^H\right)\mathclose{}.
\end{equation}
If the sample covariance matrix $\hat{\vec{R}}$ is singular, the null-spectrum of the PR-CCF estimator cannot be computed using \eqref{eq:PR-CCFPseudospectrum}. However, the diagonal loading technique \cite{RobustCaponBeamformer}, \cite{Vorobyov2014503} can be applied on the sample covariance matrix $\hat{\vec{R}}$. The choice on the loading factor $\gamma$ and its influence on the estimation performance are subject of future research.
\subsection{Partially-Relaxed Unconstrained Covariance Fitting (PR-UCF)}
Comparing with the constrained version presented in Section~\ref{subsectionPRCCF}, the formulation of the PR-UCF omits the positive semidefiniteness constraint to yield the following optimization problem:
\begin{equation}\label{eq:mainUCPBased}
\hspace*{-10pt}
\begin{aligned}
\left\lbrace\hat{\vec{a}}_{\text{PR-UCF}}\right\rbrace =\text{ }^N\underset{\vec{a}\in\mathcal{A}_1}{\arg\min}\underset{\sigma_{\text{s}}^2\geq 0,\vec{E}}{\min} &\norm{\hat{\vec{R}} - \sigma_{\text{s}}^2\vec{a}\vec{a}^H - \vec{E}\vec{E}^H}_\text{F}^2\\
\text{subject to } &\text{rank}(\vec{E})\leq N-1.
\end{aligned}
\end{equation}
When minimizing with respect to $\vec{E}$, the minimizer $\hat{\vec{E}}$ of the optimization problem in \eqref{eq:mainUCPBased} is obtained from the best rank-$(N-1)$ approximation of ${\hat{\vec{R}} - \sigma_{\text{s}}^2\vec{a}\vec{a}^H}$ as described in \eqref{eq:evd_prccf} and \eqref{eq:minimizerD_prccf}. Hence, the inner optimization of the PR-UCF estimator at each direction ${\vec{a}=\vec{a}(\vartheta)}$ is:
\begin{align}\label{eq:reformulatedUCP}
\underset{\sigma_{\text{s}}^2\geq 0}{\min}\sum\limits_{k = N}^{M} \lambda_k^2\left(\hat{\vec{R}} - \sigma_{\text{s}}^2\vec{a}\vec{a}^H\right).
\end{align}
Unlike the constrained variant in \eqref{eq:reformulatedCP}, the minimizer of $g(\sigma_{\text{s}}^2) = \sum\limits_{k = N}^{M} \lambda_k^2\left(\hat{\vec{R}} - \sigma_{\text{s}}^2\vec{a}\vec{a}^H\right)$ in \eqref{eq:reformulatedUCP} with respect to $\sigma_{\text{s}}^2$, denoted as $\hat{\sigma}_{\text{s, U}}^2$, does not have a closed form solution. However, a numerical solution of $\hat{\sigma}_{\text{s, U}}^2$ can be determined by noting that the function $g(\sigma_{\text{s}}^2)$ is continuously differentiable, and the derivative $g'\mathopen{}\left(\sigma_{\text{s}}^2\right)\mathclose{}$  is given by (see Appendix~\ref{appsec:derivEig}):
\begin{equation}\label{eq:derivUCF}
g'\mathopen{}\left(\sigma_{\text{s}}^2\right)\mathclose{} = -\sum\limits_{k=N}^{M}\dfrac{2\bar{\lambda}_k(\sigma_{\text{s}}^2)}{\sigma_{\text{s}}^4\vec{a}^H\left(\hat{\vec{R}} - \bar{\lambda}_k(\sigma_{\text{s}}^2)\vec{I}_M\right)^{-2}\vec{a}},
\end{equation} 
where we introduce the following shorthand notation:
\begin{equation}
\label{eq:shorthandLambda}
\bar{\lambda}_k(\sigma_{\text{s}}^2) = \lambda_k\left(\hat{\vec{R}} - \sigma_{\text{s}}^2\vec{a}\vec{a}^H\right).
\end{equation}
Note that the denominator in each summand of the expression in \eqref{eq:derivUCF} is always positive, we observe that:
\begin{itemize}
	\item If $\sigma_{\text{s}}^2\rightarrow 0$, then ${\bar{\lambda}_k(\sigma_{\text{s}}^2)\geq0}$ with $ k=N,\ldots,M$ and therefore:
	\begin{equation}\label{eq:zeroderiv}
		\lim\limits_{\sigma_{\text{s}}^2\rightarrow 0}{g'\mathopen{}\left(\sigma_{\text{s}}^2\right)\mathclose{}} < 0.
	\end{equation}
	\item If $\sigma_{\text{s}}^2 \rightarrow \infty$, the rank-one component $-\sigma_{\text{s}}^2\vec{a}\vec{a}^H$ is dominant to $\hat{\vec{R}}$ and thus we obtain an asymptotic result for the smallest eigenvalues $\bar{\lambda}_M\left(\sigma_{\text{s}}^2\right)$ as follows:
	\begin{align}
	\label{eq:asympEig}
	&\lim\limits_{\sigma_{\text{s}}^2\rightarrow\infty}\dfrac{\bar{\lambda}_M\left(\sigma_{\text{s}}^2\right)}{\sigma_{\text{s}}^2\norm{\vec{a}}_2^2}= -1.
	\end{align}
	In addition, the remaining eigenvalues $\bar{\lambda}_k\left(\sigma_{\text{s}}^2\right)$ with ${k=N,\ldots,(M-1)}$ are always bounded above and below thanks to the Weyl's inequality \cite{HORN199829}. Applying this remark and the identity in \eqref{eq:asympEig} to \eqref{eq:derivUCF} leads to:
	\begin{equation}\label{eq:inftyderiv}
	\lim\limits_{\sigma_{\text{s}}^2\rightarrow\infty} g'\mathopen{}\left(\sigma_{\text{s}}^2\right)\mathclose{} = \infty.
	\end{equation}
\end{itemize}
From \eqref{eq:derivUCF}, \eqref{eq:zeroderiv} and \eqref{eq:inftyderiv}, there exists a sufficiently small $\sigma_{\text{s, left}}^2$ and a sufficiently large $\sigma_{\text{s, right}}^2$ so that the sign of the derivative $g'\mathopen{}\left(\sigma_{\text{s}}^2\right)\mathclose{}$ changes from negative to positive in the interval $\left[\sigma_{\text{s, left}}^2, \sigs{right}^2\right]$. Therefore, a simple bisection search \cite{burden2010numerical} can applied to compute the minimizer $\hat{\sigma}_{\text{s, U}}^2$ of \eqref{eq:reformulatedUCP}. The steps to determine a search interval for the bisection search and the computation of the null-spectrum of the PR-UCF estimator at each direction $\vec{a}(\vartheta)$ are summarized in Algorithm~\ref{alg:PR-UCF}.
\begin{algorithm}[H]
	\caption{Calculating the null-spectrum of PR-UCF at a given direction $\vec{a} = \vec{a}(\vartheta)$}
 	\label{alg:PR-UCF}
	\begin{algorithmic}[1]
		\STATE \textbf{Initialization}:	$\sigs{left}^2 = \sigs{right}^2 >0$, tolerance $\epsilon$, the derivative $g'\mathopen{}\left(\sigs{}^2\right)\mathclose{}$ defined in \eqref{eq:derivUCF}
	 \IF{$g'\mathopen{}\left(\sigs{left}^2\right)\mathclose{}<0$}
		\REPEAT
			\STATE $\sigs{right}^2\leftarrow 2\sigs{right}^2$
		\UNTIL{$g'\mathopen{}\left(\sigs{right}^2\right)\mathclose{}>0$}
	\ELSE
		\REPEAT
			\STATE $\sigs{left}^2\leftarrow\sigs{left}^2/2$
		\UNTIL{$g'\mathopen{}\left(\sigs{left}^2\right)\mathclose{}<0$}
	\ENDIF
		\STATE Determine the root $\hat{\sigma}_{\text{s, U}}^2$ of \eqref{eq:derivUCF} by bisection search on $\left[\sigma_{\text{s, left}}^2, \sigs{right}^2\right]$  with the tolerance $\epsilon$
		\RETURN $f_{\text{PR-UCF}}(\vartheta) = \sum\limits_{k = N}^{M} \lambda_k^2\left(\hat{\vec{R}} - \hat{\sigma}_{\text{s, U}}^2\vec{a}(\vartheta)\vec{a}(\vartheta)^H\right)$
	\end{algorithmic}
\end{algorithm}
\section{Computational Aspects of the Partial Relaxation Methods}\label{sec:computationalAspect}
As introduced in Section~\ref{sec:PRMethod}, the proposed partial relaxation approach involves the estimation procedures which require extensive eigenvalue computation for the evaluation of the null-spectrum over the entire angle-of-view. In fact, the null-spectra in \eqref{eq:PR-DMLPseudospectrum}, \eqref{eq:PR-WSFPseudospectrum} and \eqref{eq:PR-CCFPseudospectrum} and Algorithm~\ref{alg:PR-UCF} depend only on the eigenvalues, and therefore the explicit computation of the eigenvectors can be avoided. Generally, if no particular structure of the matrix is exploited, the eigenvalue decomposition requires $O(K^2L)$ operations where $K$ is the dimension of the matrix and $L$ is the number of required eigenvalues \cite[Ch. 8]{golub2013matrix}. This computational complexity may be prohibitive for specific practical applications and limit the usage of the proposed partial relaxation approach in practice if no acceleration  procedure is considered. Furthermore, from an algorithmic perspective, the expressions in Equation \eqref{eq:PR-DMLPseudospectrum}, \eqref{eq:PR-WSFPseudospectrum}, \eqref{eq:PR-CCFPseudospectrum} and Algorithm \ref{alg:PR-UCF} share a common underlying problem structure in the sense that they all require, as a main task, the computation of the eigenvalues of a generic matrix form as follows:
\begin{equation}\label{eq:genericRankOneMod}
\vec{D} - \rho\vec{z}\vec{z}^H = \bar{\vec{U}}\bar{\vec{D}}\bar{\vec{U}}^H.
\end{equation}
In \eqref{eq:genericRankOneMod}, $\vec{D} = \text{diag}\left(d_1, \ldots, d_K\right)\in\realset^{K\times K}$ is a constant real diagonal matrix, $\rho\in\realset$ is an arbitrary positive real scalar and $\vec{z}= \left[z_1,\ldots,z_K\right]^T\in\complexset^{K\times 1}$ is a direction-dependent complex-valued vector. The relationship between the generic form in \eqref{eq:genericRankOneMod} and the null-spectra in \eqref{eq:PR-DMLPseudospectrum}, \eqref{eq:PR-WSFPseudospectrum}, \eqref{eq:PR-CCFPseudospectrum} and Algorithm~\ref{alg:PR-UCF} is further detailed in the sections below. Since the expression on the left hand side of \eqref{eq:genericRankOneMod} denotes a Hermitian matrix obtained by subtracting a rank-one matrix from a constant diagonal matrix, the term \textit{rank-one modified Hermitian matrix} is adopted. As presented in the following sections, this particular structure allows a faster implementation of the eigenvalue decomposition, and thus accelerates the computation of the null-spectrum of the partial relaxation estimators. 
\subsection{Eigenvalue Decomposition of a Rank-One Modified Hermitian Matrix}\label{subsec:EigTrack}
Initially proposed by Bunch, Nielsen and Sorensen in \cite{Bunch1978} as a support for calculating the eigenvalues and eigenvectors of symmetric tridiagonal matrices in parallel, the procedure of determining a complex-valued eigensystem of a rank-one modified Hermitian matrix is based on the interlacing theorem as follows \cite[p. 470]{golub2013matrix}:
\begin{theorem}
	\label{th:interlacing}
	Let $\left\lbrace d_1, \ldots, d_K\right\rbrace$ be the elements on the diagonal of the matrix $\vec{D}\in\realset^{K\times K}$ where $\left\lbrace d_1, \ldots, d_K\right\rbrace$ are distinct and sorted in descending order. Further assume that $\rho>0$ and $\vec{z}\in\complexset^{K\times 1}$ contains only non-zero entries. If the eigenvalues $\left\lbrace\bar{d}_1, \ldots, \bar{d}_K\right\rbrace$ of the matrix $\vec{D} - \rho\vec{z}\vec{z}^H$ are also sorted in descending order, then:
	\begin{itemize}
		\item $\left\lbrace\bar{d}_1, \ldots, \bar{d}_K\right\rbrace$ are the $K$ zeros of the secular function $p(x) = 0$, where $p(x)$ is given by:
		\begin{align}
		p(x) &= 1 -\rho \vec{z}^H\left(\vec{D} - x\vec{I}_K\right)^{-1}\vec{z}\\
		\label{eq:secular}
		&= 1 -\rho \sum\limits_{k=1}^{K} \dfrac{\normsca{z_k}^2}{d_k - x}.
		\end{align}
		\item $\left\lbrace\bar{d}_1, \ldots, \bar{d}_K\right\rbrace$ satisfy the interlacing property, i.e., 
		\begin{equation}
		\label{eq:interlacing}
		d_1 >\bar{d}_1 >d_2 >\bar{d}_2 >\ldots >d_K >\bar{d}_K .
		\end{equation}
		\item The eigenvector $\bar{\vec{u}}_k$ associated with the eigenvalue $\bar{d}_k$ is a multiple of $\left(\vec{D} - \bar{d}_k\vec{I}_K\right)^{-1}\vec{z}$
	\end{itemize}
\end{theorem}
The special cases of repeated elements in the diagonal matrix $\vec{D}$ and zero-valued entries in $\vec{z}$ are treated in Appendix~\ref{appsec:deflation}. Based on Theorem~\ref{th:interlacing}, rooting the secular function in \eqref{eq:secular} is of great importance for the acceleration of our proposed partial relaxation approach. Due to the structure of the secular function in \eqref{eq:secular} and the interlacing property in \eqref{eq:interlacing}, the zeros of the secular function can be determined independently of each other, thus allowing further improvement in the execution time through parallel computing. Without loss of generality, consider the $k$-th root of the secular function $\bar{d}_k$ which lies inside the interval $\left(d_{k+1}, d_{k}\right)$ where $k = 1, \ldots, K$ and $d_{K+1} = -\infty$. By defining the two auxiliary rational functions:
\begin{align}
\label{eq:exactFunc1}\psi_k(x) &\triangleq -\rho\sum\limits_{j = 1}^{k} \dfrac{\normsca{z_j}^2}{d_j - x}\\
\label{eq:exactFunc2}\phi_k(x) &\triangleq \begin{cases}
-\rho \sum\limits_{j = k+1}^{K} \dfrac{\normsca{z_j}^2}{d_j - x} &\text{ if } 1\leq k\leq K-1\\
0 &\text{ if } k = K,
\end{cases} 
\end{align}
the secular function in \eqref{eq:secular} can be rewritten as: 
\begin{align}
\label{eq:reformedSecular}
-\psi_k(x) =  1+ \phi_k(x).
\end{align}
Since both $\psi_k(x)$ and $\phi_k(x)$ are defined as the sum of multiple rational functions, a straightforward approach to solve \eqref{eq:reformedSecular} iteratively from a given point $\iter{x}{\tau}$ is using rational functions of first degree $\tilde{\psi}_k(x)$ and $\tilde{\phi}_k(x)$, respectively, as approximants. The author in \cite[Subsec. 2.2.3]{Li:CSD-94-851} suggests the approximant of type:
\vspace*{-5pt}\begin{align}
\label{eq:rationalFunc1}
R_{k;p, q}(x)  = \begin{cases}
p + \dfrac{q}{d_{k+1} - x}&\text{ if } 0\leq k\leq K-1\\
0 &\text{ if } k = K,
\end{cases}
\end{align} and choosing the parameters $p$ and $q$ such that the approximants coincide at a given point $\iter{x}{\tau}$ with the corresponding exact functions in \eqref{eq:exactFunc1} and \eqref{eq:exactFunc2}, respectively, up to the first-order derivative. A special case is obtained when $k=K$, in which $\tilde{\phi}_K(x) \triangleq 0$ is chosen. Different choices of the approximants were also introduced and discussed in \cite{Li:CSD-94-851}, \cite{MELMAN1997237}. For convenience purposes, the steps for determining the roots of the secular function in \eqref{eq:secular} are summarized in Algorithm~\ref{alg:RootingSecularFunction}.
{\setlength{\textfloatsep}{-50pt}\begin{algorithm}[t]
	\caption{Determining the $k$-th root of the secular function}
	\label{alg:RootingSecularFunction}
	\begin{algorithmic}[1]
		\STATE \textbf{Initialization}: Iteration index $\tau = 0$, arbitrary starting value $x^{(0)}\in\left(d_{k+1}, d_k\right)$, tolerance $\epsilon$
		\REPEAT
		\label{step3:Alg1}\STATE Find the parameters $p$ and $q$ such that:\vspace*{-2pt}
		\begin{align}
		\label{eq:rootStep1}R_{k-1;p,q}(x^{(\tau)}) &= \psi_k(x^{(\tau)}) \\
		\label{eq:rootStep2}R_{k-1;p,q}'(x^{(\tau)}) &= {\psi}_k'(x^{(\tau)})
		\end{align}
		\STATE Find the parameters $r$ and $s$ such that:\vspace*{-2pt}
		\begin{align}
		\label{eq:rootStep3}R_{k;r,s}(x^{(\tau)}) &= \phi_k(x^{(\tau)}) \\
		\label{eq:rootStep4}R_{k;r,s}'(x^{(\tau)}) &= \phi_k'({x^{(\tau)}})
		\end{align}
		\STATE Find $x^{(\tau+1)}\in\left(d_{k+1}, d_k\right)$ which satisfies:
		\begin{equation}
		\label{eq:rootStep5}- R_{k-1;p,q}(x^{(\tau+1)}) = 1 +R_{k;r,s}(x^{(\tau+1)})
		\end{equation}
		\STATE $\tau\leftarrow\tau +1$
		\UNTIL{$\normsca{x^{(\tau+1)}-x^{(\tau)}}<\epsilon$}
		\RETURN $\bar{d}_{k} = x^{(\tau+1)}$
	\end{algorithmic}
\end{algorithm}}
Since the approximant in \eqref{eq:rationalFunc1} is a rational function of degree one, the steps in \eqref{eq:rootStep1}-\eqref{eq:rootStep5} can be solved in closed form and thus the complexity of each iteration $\tau$ is $O(K)$. As a result, the overall complexity of the eigenvalue decomposition procedure is $O(KLI)$, where $L$ is the number of required eigenvalues, and $I$ is the number of iterations required for the convergence of Algorithm~\ref{alg:RootingSecularFunction}. Note that Algorithm~\ref{alg:RootingSecularFunction} converges quadratically \cite{Li:CSD-94-851}, and when applied to the partial relaxation methods on a fine grid, the eigenvalues for one direction can be used as a starting point for the eigenvalues at the next direction. From the simulation results, the number of iteration $I$ required for the tolerance $\epsilon = 10^{-9}$ is less than 4. Therefore, we can assume that the complexity of the computational eigenvalue decomposition using Algorithm~\ref{alg:RootingSecularFunction} is of order $O(KL)$. 
\subsection{Application to PR-DML}\label{subsec:EigTrackPRDML}
As mentioned in Section~\ref{subsec:EigTrack}, the complexity of evaluating the null-spectrum is proportional to the number of required eigenvalues $L$. Therefore, to accelerate the computation of the null-spectrum of the PR-DML method, we reduce the number of required eigenvalues by rewriting the expression in \eqref{eq:PR-DMLPseudospectrum} as follows:\vspace*{-5pt}
\begin{gather}\vspace*{-4pt}
\begin{aligned}
\hspace*{-8pt}\sum\limits_{k = N}^{M} \lambda_k\mathopen{}\left(\oproj{\vec{a}}{\hat{\vec{R}}}\right)\mathclose{} &= \tr{\oproj{\vec{a}}{\hat{\vec{R}}}} - \sum\limits_{k = 1}^{N-1}\lambda_k\mathopen{}\left(\oproj{\vec{a}}{\hat{\vec{R}}}\right)\mathclose{}\\
\label{eq:PRDML_firstreform}&= \tr{\hat{\vec{R}}} - \dfrac{\vec{a}^H\hat{\vec{R}}\vec{a}}{\vec{a}^H\vec{a}} - \sum\limits_{k = 1}^{N-1}\lambda_k\mathopen{}\left(\oproj{\vec{a}}{\hat{\vec{R}}}\right)\mathclose{}.\raisetag{2.75\baselineskip}
\end{aligned}
\end{gather}
Using the reformulation in \eqref{eq:PRDML_firstreform}, only $(N-1)$ - eigenvalues out of $M$ eigenvalues of $\oproj{\vec{a}}{\hat{\vec{R}}}$ are computed, and therefore the computational complexity is reduced. In order to apply the eigenvalue decomposition procedure presented in Section~\ref{subsec:EigTrack}, the term $\lambda_k\left(\oproj{\vec{a}}{\hat{\vec{R}}}\right)$ is further rewritten as follows:\begin{gather}
\begin{aligned}
\lambda_k\left(\oproj{\vec{a}}{\hat{\vec{R}}}\right) &= \lambda_k\left(\hat{\vec{R}}^{1/2}\oproj{\vec{a}}{\hat{\vec{R}}^{1/2}}\right)\\
&= \lambda_k\left(\hat{\vec{R}} - \dfrac{1}{\norm{\vec{a}}_2^2}\hat{\vec{R}}^{1/2}\vec{a}\vec{a}^H\hat{\vec{R}}^{1/2}\right)\\
\label{eq:PRDML_reform1}&= \lambda_k\left(\hat{\vec{\Lambda}} - \dfrac{1}{\norm{\vec{a}}_2^2}\hat{\vec{\Lambda}}^{1/2}\hat{\vec{U}}^H\vec{a}\vec{a}^H\hat{\vec{U}}\hat{\vec{\Lambda}}^{1/2}\right).\raisetag{3.5\baselineskip}
\end{aligned}
\end{gather}
From the expression in \eqref{eq:PRDML_reform1}, the eigenvalue decomposition procedure introduced in 
Section~\ref{subsec:EigTrack} is applied with ${\vec{D} = \hat{\vec{\Lambda}}}$, ${\rho = \dfrac{1}{\norm{\vec{a}}_2^2}}$ and $\vec{z} = \hat{\vec{\Lambda}}^{1/2}\hat{\vec{U}}^H\vec{a}$. From the computational perspective, except for the initial full eigenvalue decomposition in \eqref{eq:EVD}, the overall complexity of the calculation of the null-spectrum for the complete angle-of-view with $N_G$ directions is $O\left(\left(M^2 + M\left(N-1\right)\right)N_G\right) = O(M^2N_G)$. This is higher than the complexity required for computing the MUSIC null-spectrum, which is $O(MNN_G)$.
\subsection{Application to PR-WSF}
A similar iterative procedure for computing the eigenvalue decomposition as proposed in Section~\ref{subsec:EigTrack} and Section~\ref{subsec:EigTrackPRDML} can be applied directly to the PR-WSF method presented in Section~\ref{subsec:PRWSF}. However, the computational complexity of the PR-WSF method can be even further reduced due to the fact that all eigenvalues $\lambda_k(\oproj{\vec{a}}{\hat{\vec{U}}_{\text{s}}\vec{W}\hat{\vec{U}}_{\text{s}}^H})$ with $k=N+1, \ldots, M$ are equal to zero since $\text{rank}(\oproj{\vec{a}}{\hat{\vec{U}}_{\text{s}}\vec{W}\hat{\vec{U}}_{\text{s}}^H})\leq N$. Therefore, only the $N$-th eigenvalue $\lambda_N(\oproj{\vec{a}}{\hat{\vec{U}}_{\text{s}}\vec{W}\hat{\vec{U}}_{\text{s}}^H})$ needs to be calculated. Furthermore, the dimension of the matrix $\vec{D}$ in \eqref{eq:genericRankOneMod} can also be reduced. In fact, similar to \eqref{eq:PRDML_firstreform}, it can be shown that:
\begin{equation}
\label{eq:reformulatedWSF}
\begin{aligned}
&\text{ }\lambda_N\left(\oproj{\vec{a}}{\hat{\vec{U}}}_{\text{s}}\vec{W}\hat{\vec{U}}_{\text{s}}^H\right)\\
= &\text{ }\lambda_N\left(\vec{W} - \dfrac{1}{\norm{\vec{a}}_2^2}\vec{W}^{1/2}\hat{\vec{U}}_{\text{s}}^H\vec{a}\vec{a}^H\hat{\vec{U}}_{\text{s}}\vec{W}^{1/2}\right).
\end{aligned}
\end{equation}
Using the identity in \eqref{eq:reformulatedWSF}, the procedure for computing the eigenvalue decomposition introduced in Section~\ref{subsec:EigTrack} is applied with $\vec{D} = \vec{W}, \rho = \dfrac{1}{\norm{\vec{a}}_2^2}$ and $\vec{z} = \vec{W}^{1/2}\hat{\vec{U}}_{\text{s}}^H\vec{a}$. Since the dimension of the matrix is reduced from $M\times M$ to $N\times N$, and only a single eigenvalue is required, the complexity reduces to $O\left(\left(NM + N-1\right)N_G\right) = O(MNN_G)$, which is identical to the computational complexity of the MUSIC algorithm. However, the computational overhead associated with PR-WSF is still higher than MUSIC since in the preprocessing step, additional calculation for determining the weighting matrix $\vec{D} = \vec{W}$ and the vector $\vec{z} = \vec{W}^{1/2}\hat{\vec{U}}_{\text{s}}^H\vec{a}$ is required.
\subsection{Application to PR-CCF}\label{subsec:PRCCFCalc}
The expression of the PR-CCF null-spectrum in \eqref{eq:PR-CCFPseudospectrum} resembles the generic formulation of the rank-one modified Hermitian matrix in \eqref{eq:genericRankOneMod}, except for the fact that the matrix $\hat{\vec{R}}$ is generally not diagonal. Therefore, the application of the eigenvalue decomposition in Section~\ref{subsec:EigTrack} is straightforward, in which we perform an orthogonal transformation on $\hat{\vec{R}}$ and $\vec{a}$ to diagonalize $\hat{\vec{R}}$. However, the number of the eigenvalues required for the computation of the null-spectrum is $(M-N+1)$, which is typically larger than the number of sources $N$. By rewriting the expression in \eqref{eq:PR-CCFPseudospectrum} using the trace operator, only the $(N-1)$-largest eigenvalues are calculated and the null-spectrum of the PR-CCF method is rewritten in the following form to utilize the principal eigenvalues:
\hspace*{-10pt}\begin{gather}
\begin{aligned}
&\sum\limits_{k=N}^{M}\lambda_k^2\left(\hat{\vec{R}} - \hat{\sigma}_{\text{s, C}}^2\vec{a}\vec{a}^H\right) = \sum\limits_{k=N}^{M}\lambda_k\left(\left(\hat{\vec{R}} - \hat{\sigma}_{\text{s, C}}^2\vec{a}\vec{a}^H\right)^2\right)\\
=&\text{ }\tr{\left(\hat{\vec{R}} - \hat{\sigma}_{\text{s, C}}^2\vec{a}\vec{a}^H\right)^2} - \sum\limits_{k=1}^{N-1}\lambda_k^2\left(\hat{\vec{R}} - \hat{\sigma}_{\text{s, C}}^2\vec{a}\vec{a}^H\right) \\
\label{eq:PR-CCFreform1}
=&\text{ }\tr{\hat{\vec{R}}^2} - 2\hat{\sigma}_{\text{s, C}}^2\vec{a}^H\hat{\vec{R}}\vec{a}+\hat{\sigma}_{\text{s, C}}^4\norm{\vec{a}}_2^4 \\
-&\sum\limits_{k=1}^{N-1}\lambda_k^2\left(\hat{\vec{R}}-
\hat{\sigma}_{\text{s, C}}^2\vec{a}\vec{a}^H\right).\raisetag{4\baselineskip}
\end{aligned}
\end{gather}
Considering the formulation in \eqref{eq:PR-CCFreform1}, we observe that the PR-CCF method involves both the conventional and Capon beamformer in the evalutaion of the null-spectrum. Similarly to the PR-DML method, for any eigenvalue $\lambda_k\left(\hat{\vec{R}}-
\hat{\sigma}_{\text{s, C}}^2\vec{a}\vec{a}^H\right)$, it can be shown that:
\begin{equation}
\label{eq:PR-CCFreform2}
\lambda_k\left(\hat{\vec{R}}-
\hat{\sigma}_{\text{s, C}}^2\vec{a}\vec{a}^H\right) = \lambda_k\left(\hat{\vec{\Lambda}}-
\hat{\sigma}_{\text{s, C}}^2\hat{\vec{U}}^H\vec{a}\vec{a}^H\hat{\vec{U}}\right).
\end{equation}
From \eqref{eq:PR-CCFreform1} and \eqref{eq:PR-CCFreform2}, we apply the eigenvalue decomposition procedure presented in Section~\ref{subsec:EigTrack} with ${\vec{D} = \hat{\vec{\Lambda}}}$, ${\rho = \hat{\sigma}_{\text{s, C}}^2}$ and ${\vec{z} = \hat{\vec{U}}^H\vec{a}}$. Thus, the overall computational complexity of the PR-CCF algorithm is ${O\left(\left(M^2+M\left(N-1\right)\right)N_G\right) = O(M^2N_G)}$.
\subsection{Application to PR-UCF}
Unlike the PR-DML, PR-WSF and PR-CCF estimators, the PR-UCF estimator requires additional steps of calculating the derivative $g'(\sigma_{\text{s}}^2)$ in \eqref{eq:derivUCF} to obtain the minimizer $\sigma_{\text{s, U}}^2$ of \eqref{eq:reformulatedUCP}. To reduce the number of required eigenvalues for computing the derivative and the null-spectrum, the function $g(\sigma_{\text{s}}^2) = \sum\limits_{k = N}^{M} \lambda_k^2\left(\hat{\vec{R}} - \sigma_{\text{s}}^2\vec{a}\vec{a}^H\right)$ is rewritten similarly to \eqref{eq:PR-CCFreform1} as follows:
\begin{equation}
\begin{aligned}
g(\sigma_{\text{s}}^2) &=\tr{\hat{\vec{R}}^2} - 2\hat{\sigma}_{\text{s}}^2\vec{a}^H\hat{\vec{R}}\vec{a}+\hat{\sigma}_{\text{s}}^4\norm{\vec{a}}_2^4 \\
&-\sum\limits_{k=1}^{N-1}\lambda_k^2\left(\hat{\vec{R}}-
\hat{\sigma}_{\text{s}}^2\vec{a}\vec{a}^H\right).
\end{aligned}
\end{equation}
The derivative $g'\mathopen{}\left(\sigma_{\text{s}}^2\right)\mathclose{}$ is calculated as:
\begin{align}
&\begin{aligned}
g'\mathopen{}\left(\sigma_{\text{s}}^2\right)\mathclose{} &= -2\vec{a}^H\hat{\vec{R}}\vec{a} + 2\sigma_{\text{s}}^2\norm{\vec{a}}_2^4 \\&+\sum\limits_{k = 1}^{N-1}\dfrac{2\bar{\lambda}_k(\sigma_{\text{s}}^2)}{\sigma_{\text{s}}^4\vec{a}^H\left(\hat{\vec{R}} - \bar{\lambda}_k(\sigma_{\text{s}}^2)\vec{I}_M\right)^{-2}\vec{a}}
\end{aligned}
\end{align}
where $\bar{\lambda}_k(\sigma_{\text{s}}^2) = \lambda_k\left(\hat{\vec{R}} - \sigma_{\text{s}}^2\vec{a}\vec{a}^H\right)$. By substituting ${\vec{z} = \hat{\vec{U}}^H\vec{a}}$, we obtain:
\begin{align}
	\label{eq:eigPRUCFreduced}&\bar{\lambda}_k(\sigma_{\text{s}}^2) = \lambda_k\left(\hat{\vec{\Lambda}} - \sigma_{\text{s}}^2\vec{z}\vec{z}^H\right)\\
	\label{eq:derivPRUCFreduced}&\begin{aligned}
	g'\mathopen{}\left(\sigma_{\text{s}}^2\right)\mathclose{} &= -2\vec{z}^H\hat{\vec{\Lambda}}\vec{z} + 2\sigma_{\text{s}}^2\norm{\vec{z}}_2^4 \\&+\sum\limits_{k=1}^{N-1} \dfrac{2\bar{\lambda}_k\left(\sigma_{\text{s}}^2\right)}{\sigma_{\text{s}}^4\sum\limits_{j=1}^{M}\dfrac{\normsca{z_j}^2}{\left(\hat{\lambda}_j - \bar{\lambda}_k\left(\sigma_{\text{s}}^2\right)\right)^2}}.
	\end{aligned}	
\end{align}
Based on the expressions in \eqref{eq:eigPRUCFreduced} and \eqref{eq:derivPRUCFreduced}, the null-spectrum of PR-UCF from Algorithm~\ref{alg:PR-UCF} is calculated by applying the procedure in Section~\ref{subsec:EigTrack} with $\vec{D} = \hat{\vec{\Lambda}}$, $\rho = \sigma_{\text{s}, 0}^2$ and $\vec{z} = \hat{\vec{U}}^H\vec{a}$. The computational complexity of the PR-UCF method is therefore of order $O(M^2N_GN_I)$ where $N_I$ is the number of bisection steps conducted in Algorithm~\ref{alg:PR-UCF}.
\section{Simulation Results}\label{sec:SimResult}
In this section, simulation results regarding the performance of different DOA estimators are presented and compared with the stochastic Cramer-Rao Bound (CRB) \cite{MLandCRB}. The number of Monte-Carlo runs is $N_R = 1000$. The key performance indicators are the Root-Mean-Squared-Error (RMSE), which is calculated as:
\begin{equation}
\label{eq:rmseEquation}\text{RMSE} = \sqrt{\dfrac{1}{N_R N}\sum\limits_{r=1}^{N_R}\sum\limits_{n = 1}^{N}\left(\hat{\theta}^{(r)}_{n} - \theta_{\text{n}}\right)^2},
\end{equation}
and the execution time of each Monte-Carlo run. The estimated DOAs in the $r$-th Monte-Carlo run ${\hat{\vec{\theta}}^{(r)}=[\hat{\theta}^{(r)}_1,\ldots,\hat{\theta}^{(r)}_N]^T}$ and the true DOAs ${\vec{\theta}=\left[\theta_1,\ldots,\theta_N\right]^T}$ in \eqref{eq:rmseEquation} are sorted in ascending order.
The simulations are conducted in MATLAB 2016b on a PC equipped with an OS of Arch Linux with a processor 8 x Intel Core i7-6700 4.00GHz CPU and 16GB RAM. The iterative eigenvalue decomposition introduced in Section~\ref{subsec:EigTrack} is implemented in C and imported in MATLAB through a MEX interface.  In our simulations, if not further specified, we assume two uncorrelated but closely spaced source signals at $\vec{\theta} = \left[45^\circ, 50^\circ\right]^T$ which impinge on a ULA of $M = 10$ antennas with the spacing equal to half of the wavelength. We stress that unlike root-MUSIC, all PR methods are applicable to any array geometry. The source signals have the mean value of zero and unit power. The SNR is calculated as $\text{SNR} = \frac{1}{\sigma_{\text{n}}^2}$. Regarding the PR-WSF method, we choose the weighting as in \eqref{eq:optimalWeighting_WSF}. Since we consider only two source signals, the DOA estimations from the DML estimator can be obtained by performing a brute-force search for the global maximum of the objective function in \eqref{eq: DML2} over a dense grid on $\mathcal{A}_2$. As depicted in Figure~\ref{fig:uncorrSNR}, the partial relaxation methods exhibit superior SNR threshold performance in comparison to the MUSIC algorithm. The PR-CCF and PR-UCF estimator possess almost identical estimation error performance in the inspected SNR region, where their thresholds occur at an even lower SNR than that of root-MUSIC. The PR-CCF and PR-UCF are outperformed by the brute-force DML in both the asymptotic and the non-asymptotic regions, although the difference in RMSE is small. This remark suggests that PR-CCF is more favorable than PR-UCF, since the computational complexity of PR-CCF is lower than that of PR-UCF while the error performances are comparable. The RMSE of both PR-CCF and PR-UCF does not approach the CRB. However, the difference in RMSE is insignificant. PR-DML and PR-WSF have similar performance behaviors, achieving the CRB at a lower SNR than MUSIC but much higher than root-MUSIC.\\
{\setlength{\textfloatsep}{-20pt}\setlength{\intextsep}{-50pt}\begin{figure}[t]
	\centering
%
%
\definecolor{mycolor1}{RGB}{136, 0, 21}%
\definecolor{mycolor2}{RGB}{185,122,187}%
\definecolor{mycolor3}{RGB}{255,32,27}%
\definecolor{mycolor4}{RGB}{255,174,201}%
\definecolor{mycolor5}{RGB}{255,137,29}%
\definecolor{mycolor6}{RGB}{255,201,14}%
\definecolor{mycolor7}{RGB}{255,242,0}%
\definecolor{mycolor8}{RGB}{239,228,176}%
\definecolor{mycolor9}{RGB}{31,194,43}%
\definecolor{mycolor10}{RGB}{181,230,29}%
\definecolor{mycolor11}{RGB}{0,119,251}%
\definecolor{mycolor12}{RGB}{94,243,255}%

\definecolor{mycolor13}{RGB}{63,72,204}%
\definecolor{mycolor14}{RGB}{112,146,190}%
\definecolor{mycolor15}{RGB}{163,73,164}%
\definecolor{mycolor16}{RGB}{200,191,231}%
\begin{tikzpicture}

\begin{axis}[%
width=6.5cm,
height=4.5cm,
at={(0in,0in)},
scale only axis,
every tick label/.append style={font=\scriptsize},
xmin=-10,
xmax=20,
xlabel={\small SNR (dB)},
ymode=log,
ymin=0.05,
ymax=100,
ylabel={\small RMSE (deg)},
y label style={at={(axis description cs:-0.08,.5)},anchor=south},
yminorticks=true,
grid = both,
axis background/.style={fill=white},
legend columns = 2,
legend style={legend cell align=left, align=left, draw=white!15!black, row sep=-2pt, at={(0.02,0.02)}, anchor = south west}
]
\addplot [color=mycolor12, line width = 1pt, mark = diamond, mark options={solid, mycolor12}, mark repeat = 2]
  table[row sep=crcr]{%
-10	47.3031647864437\\
-9	47.7271604565532\\
-8	47.3991796831618\\
-7	48.806565999172\\
-6	48.0288997770879\\
-5	48.3552808997977\\
-4	49.3904197992694\\
-3	48.8707907370013\\
-2	50.5874872442596\\
-1	51.3745748561046\\
0	51.3324829674022\\
1	52.3351794093495\\
2	51.6804003809464\\
3	53.0348361802463\\
4	53.5911340259021\\
5	52.6598248152442\\
6	54.0481389601122\\
7	49.8096039667984\\
8	47.2714443572658\\
9	43.153638218975\\
10	35.080211473116\\
11	26.6791590879392\\
12	19.8948309076129\\
13	13.2424437802639\\
14	5.58865061024093\\
15	3.43525136042935\\
16	0.291821536149141\\
17	0.262404022719196\\
18	0.217988222295274\\
19	0.199581057694887\\
20	0.182092889525796\\
};
\addlegendentry{\tiny MUSIC}

\addplot [color=mycolor11, line width = 1pt, mark = o, mark options={solid, mycolor11}, mark repeat = 2]
  table[row sep=crcr]{%
-10	46.6836871349844\\
-9	48.0951589877489\\
-8	47.3734592667006\\
-7	48.7497348005865\\
-6	49.220316412109\\
-5	48.1830797316872\\
-4	47.519366167965\\
-3	43.844025550346\\
-2	42.8779198819604\\
-1	35.3263589020957\\
0	29.4848154174065\\
1	18.8171101735619\\
2	10.0219294753479\\
3	4.95458340462117\\
4	1.27118645720794\\
5	1.23320114870736\\
6	0.918899150129384\\
7	0.765285568897414\\
8	0.643453948430755\\
9	0.535784762580989\\
10	0.486743725897028\\
11	0.414928342858068\\
12	0.371283540662969\\
13	0.335254534259638\\
14	0.288597894864492\\
15	0.25802187663195\\
16	0.230182922577899\\
17	0.200330475389217\\
18	0.173713626013142\\
19	0.161907763586291\\
20	0.14882481027506\\
};
\addlegendentry{\tiny root-MUSIC}

\addplot [color=mycolor9, line width = 1pt, mark = square, mark options={solid, mycolor9}, mark repeat = 2]
  table[row sep=crcr]{%
-10	44.2876960942924\\
-9	45.9160868952541\\
-8	45.4521669092871\\
-7	44.4783999278729\\
-6	44.3608016312791\\
-5	44.1186504455003\\
-4	41.47235579229\\
-3	41.570680023232\\
-2	42.4336739582065\\
-1	40.4119576599332\\
0	40.5522225271362\\
1	39.3263793522893\\
2	37.6772935113862\\
3	33.6234914836527\\
4	33.2362619927127\\
5	28.9379795404488\\
6	27.8097734986043\\
7	21.6819541004394\\
8	18.4952021656484\\
9	13.3220190013954\\
10	9.36658744260813\\
11	6.89334348124793\\
12	3.45142946369184\\
13	0.660225182842023\\
14	0.308595476811692\\
15	0.277839866646945\\
16	0.247279138631172\\
17	0.216658421826954\\
18	0.193297543733231\\
19	0.178295930020058\\
20	0.165951182796574\\
};
\addlegendentry{\tiny PR-DML}

\addplot [color=mycolor6, mark=asterisk, mark options={solid, mycolor6}, line width = 1pt, dashdotted, mark repeat = 2]
table[row sep=crcr]{%
-10	44.9752011784797\\
-9	44.8586561162214\\
-8	44.0144946442511\\
-7	44.2025406448228\\
-6	42.2074006963712\\
-5	41.6050289108739\\
-4	38.9258550630523\\
-3	37.7988242756563\\
-2	36.1442370618532\\
-1	36.0401437494718\\
0	33.4660796358255\\
1	32.2326370177097\\
2	31.8360309537696\\
3	29.3288399582142\\
4	29.595644243535\\
5	24.9863630007651\\
6	24.9742236850204\\
7	19.0956469872063\\
8	16.9021027820977\\
9	10.8115276232126\\
10	8.87378835499188\\
11	6.04657774957194\\
12	3.44846022411303\\
13	0.654317157179446\\
14	0.308245504896501\\
15	0.276819769513374\\
16	0.247859290957431\\
17	0.216602166509572\\
18	0.193509081873578\\
19	0.178656845645713\\
20	0.165903967939932\\
};
\addlegendentry{\tiny PR-WSF}

\addplot [color=mycolor3, line width = 1pt, mark = triangle, mark options={solid, mycolor3}, mark repeat = 2, dashdotted]
  table[row sep=crcr]{%
-10	46.4547044804078\\
-9	45.9207995075701\\
-8	44.3793583627083\\
-7	42.275327620208\\
-6	37.4737469286386\\
-5	36.4048190988966\\
-4	28.0364785043771\\
-3	21.1598916742633\\
-2	14.0519381718767\\
-1	7.93199562743924\\
0	4.33303410884251\\
1	2.8310965913395\\
2	1.87580390347093\\
3	1.66290838721516\\
4	1.44615916553599\\
5	1.24700697457592\\
6	1.09928878801434\\
7	0.946713907816305\\
8	0.834356618539575\\
9	0.705701185732499\\
10	0.627452640912216\\
11	0.543079019114137\\
12	0.469194926790052\\
13	0.407019075578877\\
14	0.372558754472153\\
15	0.333018622032334\\
16	0.283865084266935\\
17	0.257292034547305\\
18	0.22612621297924\\
19	0.206171501729531\\
20	0.184557227439277\\
};
\addlegendentry{\tiny PR-CCF}

\addplot [color=mycolor3, line width = 1pt,  mark repeat = 2]
table[row sep=crcr]{%
	-10	46.3817217383385\\
	-9	45.6865481699804\\
	-8	42.6760965332818\\
	-7	42.4664126685567\\
	-6	38.7533534849875\\
	-5	34.5004352429935\\
	-4	27.5931132823293\\
	-3	22.7481561250639\\
	-2	15.2006421414128\\
	-1	7.89444030614192\\
	0	4.66128575246892\\
	1	3.3319861767854\\
	2	1.89303209713772\\
	3	1.59571482858676\\
	4	1.376601234847\\
	5	1.21503697381402\\
	6	1.04853796881027\\
	7	0.927583282865787\\
	8	0.797898508012917\\
	9	0.679969040081881\\
	10	0.594448985654758\\
	11	0.519629630383092\\
	12	0.443666098385064\\
	13	0.384713393210676\\
	14	0.339575047716326\\
	15	0.309555859438827\\
	16	0.266008390009185\\
	17	0.233198767480931\\
	18	0.208814469472707\\
	19	0.188699041365821\\
	20	0.171228033746523\\
};
\addlegendentry{\tiny PR-UCF}

\addplot [color=mycolor9, line width = 1pt, dashdotted]
table[row sep=crcr]{%
	-10	40.0274670650099\\
	-9	37.2359152397805\\
	-8	36.2233989429649\\
	-7	31.6541401434047\\
	-6	27.6363947579765\\
	-5	24.3811880041419\\
	-4	17.886076708354\\
	-3	12.9079770930873\\
	-2	8.67251431161417\\
	-1	3.93806645137197\\
	0	1.82368589130554\\
	1	1.57486824364179\\
	2	1.3636729716792\\
	3	1.19251973388494\\
	4	1.04005294992252\\
	5	0.90693486422732\\
	6	0.780443557940843\\
	7	0.732457773607615\\
	8	0.59913209787836\\
	9	0.543395659248514\\
	10	0.476012589683723\\
	11	0.426172239554359\\
	12	0.363882334949639\\
	13	0.325599753588328\\
	14	0.295169878778142\\
	15	0.260763440421077\\
	16	0.23082574386328\\
	17	0.207416220799079\\
	18	0.189205673785303\\
	19	0.168541225005849\\
	20	0.146425100037656\\
};
\addlegendentry{\tiny DML}

\addplot [color=black, line width = 1pt]
  table[row sep=crcr]{%
-10	8.10262977040441\\
-9	6.79921193761549\\
-8	5.72557924439238\\
-7	4.83781487950733\\
-6	4.10135811147125\\
-5	3.48876500193846\\
-4	2.97799873023273\\
-3	2.5511655370097\\
-2	2.19360303397964\\
-1	1.89322785467855\\
0	1.64006124179588\\
1	1.42587010835279\\
2	1.24388130069183\\
3	1.088543659468\\
4	0.955324222253411\\
5	0.840531891898854\\
6	0.741165393717943\\
7	0.65478369276096\\
8	0.57939725931843\\
9	0.513378327967723\\
10	0.455387984889173\\
11	0.404317729315413\\
12	0.359243142943292\\
13	0.319387444720174\\
14	0.28409295498893\\
15	0.252798787464721\\
16	0.225023386641749\\
17	0.200350804286469\\
18	0.178419847676086\\
19	0.158915430021826\\
20	0.141561611883994\\
};
\addlegendentry{\tiny CRB}

\end{axis}
\end{tikzpicture}%
	\caption{Uncorrelated source signals, number of snapshots $T =40$}
	\label{fig:uncorrSNR}
\end{figure}
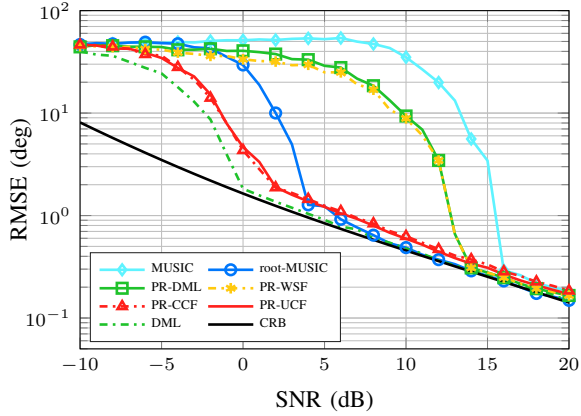
}
In the next simulation, we consider the correlated signals of two sources. The correlation coefficient $\rho$ is defined as:
\begin{equation}
\rho =\dfrac{\EV{s_1(t)^Hs_2(t)}}{\sqrt{\EV{\normsca{s_1(t)}^2}\EV{\normsca{s_2(t)}^2}}}.
\end{equation} In Figure~\ref{fig:corrSNR}, the correlation coefficient is set to $\rho = 0.95$. The number of snapshots is increased to $T = 200$. Note that spatial smoothing \cite{SpatialSmoothing, SpatialSmoothingArticle} or forward-backward averaging \cite{FBA2} is not applied. DML consistently outperforms other considered estimators in the inspected SNR region. The threshold of root-MUSIC now occurs at a slightly lower SNR than that of the partial relaxation methods. On the other hand, in the post-threshold region, all estimators under the partial relaxation framework have a lower RMSE than \text{root-MUSIC}. However, the improvement in RMSE is negligible.\\
{\setlength{\textfloatsep}{-40pt}\setlength{\intextsep}{-50pt}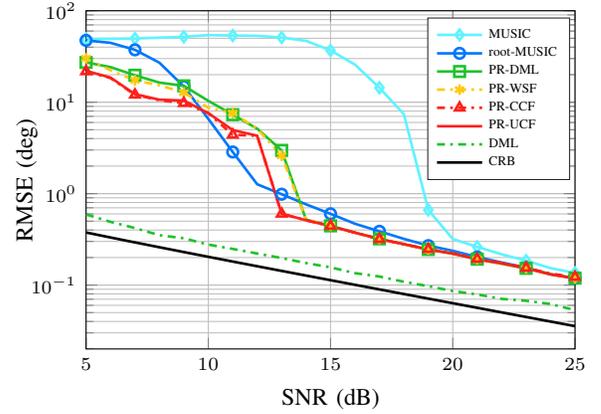
\begin{figure}[t]
	\centering
%
%
\definecolor{mycolor1}{RGB}{136, 0, 21}%
\definecolor{mycolor2}{RGB}{185,122,187}%
\definecolor{mycolor3}{RGB}{255,32,27}%
\definecolor{mycolor4}{RGB}{255,174,201}%
\definecolor{mycolor5}{RGB}{255,137,29}%
\definecolor{mycolor6}{RGB}{255,201,14}%
\definecolor{mycolor7}{RGB}{255,242,0}%
\definecolor{mycolor8}{RGB}{239,228,176}%
\definecolor{mycolor9}{RGB}{31,194,43}%
\definecolor{mycolor10}{RGB}{181,230,29}%
\definecolor{mycolor11}{RGB}{0,119,251}%
\definecolor{mycolor12}{RGB}{94,243,255}%

\definecolor{mycolor13}{RGB}{63,72,204}%
\definecolor{mycolor14}{RGB}{112,146,190}%
\definecolor{mycolor15}{RGB}{163,73,164}%
\definecolor{mycolor16}{RGB}{200,191,231}%
\begin{tikzpicture}

\begin{axis}[%
width=6.5cm,
height=4.5cm,
at={(0in,0in)},
scale only axis,
every tick label/.append style={font=\scriptsize},
xmin=5,
xmax=25,
xlabel={\small SNR (dB)},
ymode=log,
ymin=0.02,
ymax=100,
ylabel={\small RMSE (deg)},
y label style={at={(axis description cs:-0.08,.5)},anchor=south},
yminorticks=true,
grid = both,
axis background/.style={fill=white},
legend style={legend cell align=left, align=left, draw=white!15!black, row sep=-2pt, at={(0.99,0.98)}, anchor = north east}
]
\addplot [color=mycolor12, line width = 1pt, mark = diamond, mark options={solid, mycolor12}, mark repeat = 2]
table[row sep=crcr]{%
5	49.1992157703541\\
6	49.2712420819671\\
7	49.6829788726667\\
8	50.7875547769463\\
9	51.8129225993126\\
10	54.1719749904477\\
11	53.5243634495457\\
12	53.2115530357566\\
13	50.961836645424\\
14	46.8463120933674\\
15	36.8246633819304\\
16	25.7723538072408\\
17	14.3731840110969\\
18	7.38175467608595\\
19	0.662593708837826\\
20	0.318931656504981\\
21	0.261867816373351\\
22	0.216996022534516\\
23	0.184799030232837\\
24	0.153165398416688\\
25	0.13574135052539\\};
\addlegendentry{\tiny MUSIC}

\addplot [color=mycolor11, line width = 1pt, mark = o, mark options={solid, mycolor11}, mark repeat = 2]
table[row sep=crcr]{%
5	47.3981113407874\\
6	44.587281497453\\
7	37.3679975433954\\
8	27.0282009708238\\
9	14.8413339771145\\
10	6.56787683245051\\
11	2.85340723611239\\
12	1.26664167168022\\
13	0.98487662499692\\
14	0.763664518110812\\
15	0.60135828866949\\
16	0.46829277402403\\
17	0.387213606379161\\
18	0.320029894960711\\
19	0.269961375388414\\
20	0.237013773846307\\
21	0.202650486179545\\
22	0.176596508633908\\
23	0.155174033602878\\
24	0.131380398702991\\
25	0.118748801731618\\
};
\addlegendentry{\tiny root-MUSIC}

\addplot [color=mycolor9, line width = 1pt, mark = square, mark options={solid, mycolor9}, mark repeat = 2]
table[row sep=crcr]{%
5	27.2463207149773\\
6	24.1835013836421\\
7	19.7051622828337\\
8	16.3697720398471\\
9	15.086454266513\\
10	10.3435255392911\\
11	7.29856376345357\\
12	5.13148828636196\\
13	2.97108310871897\\
14	0.518663516850357\\
15	0.444361745445422\\
16	0.374444613979278\\
17	0.320325879880572\\
18	0.27943949247955\\
19	0.245117405145946\\
20	0.22062886331811\\
21	0.191936191627104\\
22	0.171275604432002\\
23	0.15242557135738\\
24	0.130817220297371\\
25	0.119071156243068\\
};
\addlegendentry{\tiny PR-DML}

\addplot [color=mycolor6, mark=asterisk, mark options={solid, mycolor6}, line width = 1pt, dashdotted, mark repeat = 2]
table[row sep=crcr]{%
5	30.2195754398628\\
6	22.3097835578244\\
7	17.3362816931551\\
8	15.2990652983976\\
9	12.7335896019713\\
10	8.69366325797012\\
11	7.61310976851591\\
12	5.12280851242846\\
13	2.59660257607828\\
14	0.519650452820798\\
15	0.445210161673197\\
16	0.375000086258359\\
17	0.319823268440252\\
18	0.278783908951189\\
19	0.245485842047207\\
20	0.220688095457577\\
21	0.19159613515083\\
22	0.170621660363281\\
23	0.152347101545744\\
24	0.130911782665262\\
25	0.119208869120399\\
};
\addlegendentry{\tiny PR-WSF}

\addplot [color=mycolor3, line width = 1pt, mark = triangle, mark options={solid, mycolor3}, mark repeat = 2, dashdotted]
table[row sep=crcr]{%
	5	21.7244902023957\\
	6	18.3679453857\\
	7	12.0608948096656\\
	8	10.4979851533125\\
	9	9.76306953221751\\
	10	7.59090811081034\\
	11	4.3930955486749\\
	12	4.2936133082363\\
	13	0.601556538371325\\
	14	0.509811302335009\\
	15	0.441886966534041\\
	16	0.374370711580564\\
	17	0.31778474775676\\
	18	0.280014103494821\\
	19	0.246645357765211\\
	20	0.222318300140225\\
	21	0.193084591495483\\
	22	0.175282907150732\\
	23	0.154240092857181\\
	24	0.133353124941166\\
	25	0.122323255919072\\
};
\addlegendentry{\tiny PR-CCF}

\addplot [color=mycolor3, line width = 1pt,  mark repeat = 2]
table[row sep=crcr]{%
5	22.0101370783772\\
6	18.5966709708371\\
7	12.2409714907061\\
8	10.6449734201861\\
9	10.4167477711543\\
10	7.59201685481316\\
11	4.95392908805979\\
12	4.29765396871735\\
13	0.603531847389726\\
14	0.511579376900496\\
15	0.442081382506809\\
16	0.373925131227417\\
17	0.318903317460006\\
18	0.280160053610591\\
19	0.24496192878883\\
20	0.221092214072364\\
21	0.191477807790773\\
22	0.172474511534537\\
23	0.152627946477537\\
24	0.131009460783923\\
25	0.118199169883993\\
};
\addlegendentry{\tiny PR-UCF}

\addplot [color=mycolor9, line width = 1pt, dashdotted]
  table[row sep=crcr]{%
5	0.592796604075166\\
6	0.491145349577038\\
7	0.420192065276396\\
8	0.351018600415568\\
9	0.324575007414085\\
10	0.277164754823166\\
11	0.248507681830045\\
12	0.221295253270689\\
13	0.197519972335947\\
14	0.173969119449177\\
15	0.156299730012646\\
16	0.13504301278481\\
17	0.123947035383577\\
18	0.108680369094671\\
19	0.0972406827642396\\
20	0.0860692749705547\\
21	0.0789264769023352\\
22	0.070612061453393\\
23	0.0673114799436969\\
24	0.0620243516137834\\
25	0.0532322490750898\\
};
\addlegendentry{\tiny DML}

\addplot [color=black, line width = 1pt]
table[row sep=crcr]{%
5	0.375897289508469\\
6	0.331459240584743\\
7	0.292828169514369\\
8	0.259114331562617\\
9	0.229589767902202\\
10	0.203655698069767\\
11	0.180816385451524\\
12	0.160658417614375\\
13	0.142834407510853\\
14	0.127050231856807\\
15	0.113055054680128\\
16	0.100633517811634\\
17	0.0895996035462602\\
18	0.0797917815877772\\
19	0.0710691408404829\\
20	0.0633082774354105\\
21	0.056400765726455\\
22	0.0502510814756979\\
23	0.0447748786334965\\
24	0.039897545281621\\
25	0.0355529823385494\\
};
\addlegendentry{\tiny CRB}

\end{axis}
\end{tikzpicture}%
	\caption{Correlated source signals with $\rho = 0.95$, number of snapshots $T = 200$}
	\label{fig:corrSNR}
\end{figure}
}
In the next simulation, as depicted in Figure~\ref{fig:uncorrT}, the SNR is fixed at $3$ dB, and the number of snapshots $T$ is varied between 10 and 10000. The RMSE performance of PR-UCF/PR-CCF resembles that of DML, achieving the asymptotic region at $T=30$ samples, which is approximately an order of magnitude lower in the required snapshots than that for PR-DML/PR-WSF. However, in the post-threshold region, the RMSE of PR-CCF and PR-UCF is not as close to the CRB as that of root-MUSIC, PR-DML or PR-WSF. PR-WSF outperforms PR-DML consistently in this simulation. In general, the partial relaxation methods outperform MUSIC.\\
{\setlength{\textfloatsep}{-20pt}\begin{figure}[t]
	\centering
%
%
\definecolor{mycolor1}{RGB}{136, 0, 21}%
\definecolor{mycolor2}{RGB}{185,122,187}%
\definecolor{mycolor3}{RGB}{255,32,27}%
\definecolor{mycolor4}{RGB}{255,174,201}%
\definecolor{mycolor5}{RGB}{255,137,29}%
\definecolor{mycolor6}{RGB}{255,201,14}%
\definecolor{mycolor7}{RGB}{255,242,0}%
\definecolor{mycolor8}{RGB}{239,228,176}%
\definecolor{mycolor9}{RGB}{31,194,43}%
\definecolor{mycolor10}{RGB}{181,230,29}%
\definecolor{mycolor11}{RGB}{0,119,251}%
\definecolor{mycolor12}{RGB}{94,243,255}%

\definecolor{mycolor13}{RGB}{63,72,204}%
\definecolor{mycolor14}{RGB}{112,146,190}%
\definecolor{mycolor15}{RGB}{163,73,164}%
\definecolor{mycolor16}{RGB}{200,191,231}%
\begin{tikzpicture}

\begin{axis}[%
width=6.5cm,
height=4.5cm,
at={(0in,0in)},
scale only axis,
every tick label/.append style={font=\scriptsize},
xmode=log,
xmin=10,
xmax=10000,
xlabel={\small Number of Snapshots},
ymode=log,
ymin=0.02,
ymax=100,
ylabel={\small RMSE (deg)},
y label style={at={(axis description cs:-0.08,.5)},anchor=south},
yminorticks=true,
grid = both,
axis background/.style={fill=white},
legend columns=2, 
legend style={legend cell align=left, align=left, draw=white!15!black, row sep=-2pt, at={(0.02,0.02)}, anchor = south west}
]
\addplot [color=mycolor12, line width = 1pt, mark = diamond, mark options={solid, mycolor12}, mark repeat = 2]
  table[row sep=crcr]{%
10	50.1946300314499\\
14	51.5745528090874\\
21	51.3138932699903\\
30	53.5936063678806\\
43	54.0289570315439\\
62	52.7008829579101\\
89	52.5421154743332\\
127	49.7870049342205\\
183	41.3493899748033\\
264	28.7678175379994\\
379	16.2464010865188\\
546	5.30537349139166\\
785	0.339750337560946\\
1129	0.258064455907273\\
1624	0.203371362943564\\
2336	0.163499004962934\\
3360	0.130596205728571\\
4833	0.108076721891395\\
6952	0.0875095803458799\\
10000	0.0721228199740362\\
};
\addlegendentry{\tiny MUSIC}

\addplot [color=mycolor11, line width = 1pt, mark = o, mark options={solid, mycolor11}, mark repeat = 2]
  table[row sep=crcr]{%
10	39.623683408757\\
14	32.3027950715943\\
21	23.6455333486277\\
30	13.0376315785364\\
43	4.69738878517297\\
62	1.14555664244075\\
89	0.934099956926485\\
127	0.724947578623831\\
183	0.559934408833335\\
264	0.448001213885763\\
379	0.357660764108104\\
546	0.307931768717933\\
785	0.250064905597418\\
1129	0.208959739038993\\
1624	0.172915410636016\\
2336	0.14504925764739\\
3360	0.119763875806759\\
4833	0.100443248894979\\
6952	0.0834156808889878\\
10000	0.0690235775131416\\
};
\addlegendentry{\tiny root-MUSIC}

\addplot [color=mycolor9, line width = 1pt, mark = square, mark options={solid, mycolor9}, mark repeat = 2]
  table[row sep=crcr]{%
10	35.9164858040519\\
14	36.782765494868\\
21	36.9315657498483\\
30	36.1710204243165\\
43	33.6442368264743\\
62	31.6865040251034\\
89	28.1345198246963\\
127	22.5966158385722\\
183	13.3542111969941\\
264	5.70498998204408\\
379	0.662959197238068\\
546	0.322839676492553\\
785	0.2588635549476\\
1129	0.213929012599459\\
1624	0.176325846717505\\
2336	0.147116736413304\\
3360	0.122175672548662\\
4833	0.10191191923499\\
6952	0.0848534200646184\\
10000	0.0706270470621108\\
};
\addlegendentry{\tiny PR-DML}

\addplot [color=mycolor6, mark=asterisk, mark options={solid, mycolor6}, line width = 1pt, dashdotted, mark repeat = 2]
table[row sep=crcr]{%
10	33.4965446437579\\
14	33.4520754663935\\
21	33.4373790217215\\
30	34.2270949671983\\
43	28.5497080909072\\
62	26.7634283984722\\
89	22.6053219977734\\
127	16.4435536087713\\
183	9.23508359342856\\
264	3.16042550653398\\
379	0.367254239862883\\
546	0.315080014298192\\
785	0.25347553797805\\
1129	0.211708231169956\\
1624	0.174445212693132\\
2336	0.146667129488489\\
3360	0.121066191946215\\
4833	0.101663784660604\\
6952	0.0848435931799453\\
10000	0.0707287972904495\\
};
\addlegendentry{\tiny PR-WSF}

\addplot [color=mycolor3, line width = 1pt, mark = triangle, mark options={solid, mycolor3}, mark repeat = 2, dashdotted]
table[row sep=crcr]{%
	10	14.755615339877\\
	14	9.65439465640243\\
	21	5.04719265900314\\
	30	2.12673592345484\\
	43	1.51342666805998\\
	62	1.23088128306915\\
	89	0.975382467700222\\
	127	0.796766506925852\\
	183	0.64707461724881\\
	264	0.504400218827499\\
	379	0.400419085519123\\
	546	0.334161687107826\\
	785	0.266656725903018\\
	1129	0.219907605281793\\
	1624	0.178676899755966\\
	2336	0.149277690237831\\
	3360	0.122486931447418\\
	4833	0.102331382632268\\
	6952	0.0857623987751428\\
	10000	0.070255669145002\\
};
\addlegendentry{\tiny PR-CCF}

\addplot [color=mycolor3, line width = 1pt,  mark repeat = 2]
  table[row sep=crcr]{%
10	32.2452593249025\\
14	12.4206403231216\\
21	4.36386851757579\\
30	1.93295506823594\\
43	1.54930878592183\\
62	1.25149513552835\\
89	0.986436144870433\\
127	0.802012138050325\\
183	0.649197585947858\\
264	0.505132812475304\\
379	0.401421157802454\\
546	0.334560782953345\\
785	0.266857810986056\\
1129	0.219804734710675\\
1624	0.178622643407906\\
2336	0.149264191239483\\
3360	0.122479840379489\\
4833	0.102331382632268\\
6952	0.0857623987751428\\
10000	0.070255669145002\\
};
\addlegendentry{\tiny PR-UCF}

\addplot [color=mycolor9, line width = 1pt, dashdotted]
table[row sep=crcr]{%
10	7.30224211690923\\
14	3.316350339482\\
21	2.0824181157874\\
30	1.37019020936219\\
43	1.12038219834009\\
62	0.947795477385608\\
89	0.791552062636527\\
127	0.641508867810451\\
183	0.516832579335939\\
264	0.448885999446976\\
379	0.356434058308786\\
546	0.302610944787772\\
785	0.24839592733166\\
1129	0.207913600639375\\
1624	0.16916053478427\\
2336	0.147027075070798\\
3360	0.120443687559946\\
4833	0.0998290761271472\\
6952	0.0861330888748355\\
10000	0.0703991096692896\\
};
\addlegendentry{\tiny DML}

\addplot [color=black, line width = 1pt]
  table[row sep=crcr]{%
10	2.17708731893599\\
14	1.83997461051425\\
21	1.50233297847871\\
30	1.25694194963702\\
43	1.04988473279511\\
62	0.874339307638954\\
89	0.729761327326964\\
127	0.610905433244677\\
183	0.508920919612497\\
264	0.423715044784778\\
379	0.35363581561591\\
546	0.294631737429882\\
785	0.245720290317115\\
1129	0.204893828374265\\
1624	0.170837353768317\\
2336	0.142442448037361\\
3360	0.118769850399436\\
4833	0.0990301531228937\\
6952	0.0825697493822608\\
10000	0.0688455459290716\\
};
\addlegendentry{\tiny CRB}

\end{axis}

\end{tikzpicture}%
	\caption{Uncorrelated source signals, $\text{SNR} = 3 \text{ dB}$}
	\label{fig:uncorrT}
\end{figure}
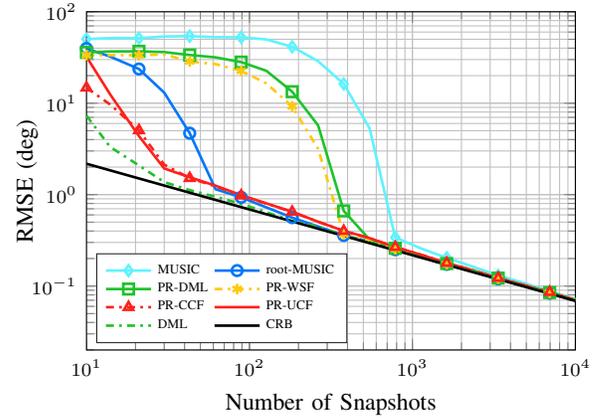
}
In the fourth setup, the DOA of the first source signal is fixed at $\theta_1 = 45^\circ$ and the angular separation between two sources $\Delta\theta$ is varied from $0.5^\circ$ to $6^\circ$. In Figure~\ref{fig:uncorrDelta}, similar to DML, the RMSE of PR-CCF/PR-UCF is close to the Cramer-Rao Bound even when the angular separation $\Delta\theta$ is as small as $1.25^\circ$, which is significantly smaller than the angular separation required for PR-DML/PR-WSF to resolve two sources. However, the RMSE of PR-CCF/PR-UCF slowly achieves the CRB only when $\Delta\theta>5^\circ$. PR-WSF slightly outperforms PR-DML, and both algorithms outperform MUSIC.\\
{\setlength{\textfloatsep}{-20pt}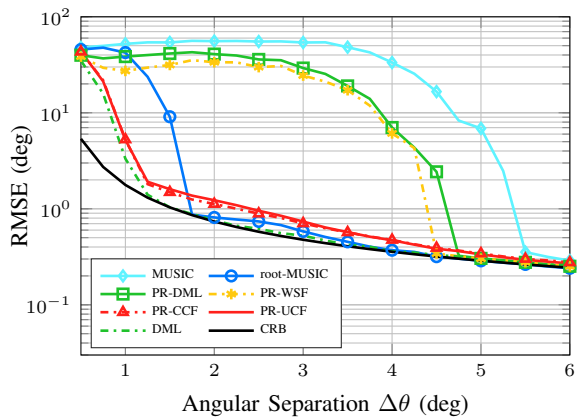
\begin{figure}[t]
	\centering
%
%
\definecolor{mycolor1}{RGB}{136, 0, 21}%
\definecolor{mycolor2}{RGB}{185,122,187}%
\definecolor{mycolor3}{RGB}{255,32,27}%
\definecolor{mycolor4}{RGB}{255,174,201}%
\definecolor{mycolor5}{RGB}{255,137,29}%
\definecolor{mycolor6}{RGB}{255,201,14}%
\definecolor{mycolor7}{RGB}{255,242,0}%
\definecolor{mycolor8}{RGB}{239,228,176}%
\definecolor{mycolor9}{RGB}{31,194,43}%
\definecolor{mycolor10}{RGB}{181,230,29}%
\definecolor{mycolor11}{RGB}{0,119,251}%
\definecolor{mycolor12}{RGB}{94,243,255}%

\definecolor{mycolor13}{RGB}{63,72,204}%
\definecolor{mycolor14}{RGB}{112,146,190}%
\definecolor{mycolor15}{RGB}{163,73,164}%
\definecolor{mycolor16}{RGB}{200,191,231}%
\begin{tikzpicture}

\begin{axis}[%
width=6.5cm,
height=4.5cm,
at={(0in,0in)},
scale only axis,
every tick label/.append style={font=\scriptsize},
xmin=0.5,
xmax=6,
xlabel={\small Angular Separation $\Delta\theta$ (deg)},
ymode=log,
ymin=0.03,
ymax=100,
ylabel={\small RMSE (deg)},
y label style={at={(axis description cs:-0.08,.5)},anchor=south},
grid = both,
yminorticks=true,
axis background/.style={fill=white},
legend columns=2, 
legend style={legend cell align=left, align=left, draw=white!15!black, row sep=-2pt, at={(0.02,0.02)}, anchor = south west}
]
\addplot [color=mycolor12, line width = 1pt, mark = diamond, mark options={solid, mycolor12}, mark repeat = 2]
  table[row sep=crcr]{%
0.5	47.9488521904622\\
0.75	49.7777477450816\\
1	52.2456556740559\\
1.25	54.1340503603877\\
1.5	54.0613857741896\\
1.75	56.1766695954269\\
2	55.7330098684993\\
2.25	55.9918384531807\\
2.5	55.23831930835\\
2.75	55.4240757378131\\
3	53.9767010117379\\
3.25	54.1969130970462\\
3.5	48.2052136647552\\
3.75	42.5687675160735\\
4	33.3691750597757\\
4.25	25.5263109674489\\
4.5	16.619880856565\\
4.75	8.29815949313582\\
5	6.87629375166583\\
5.25	2.47892064994397\\
5.5	0.353988515291587\\
5.75	0.315849611126216\\
6	0.288567438313612\\
};
\addlegendentry{\tiny MUSIC}

\addplot [color=mycolor11, line width = 1pt, mark = o, mark options={solid, mycolor11}, mark repeat = 2]
  table[row sep=crcr]{%
0.5	45.6112552282805\\
0.75	47.6551964064095\\
1	42.3570510917722\\
1.25	23.7194676802574\\
1.5	9.09973635633609\\
1.75	0.869552739760562\\
2	0.812808039156334\\
2.25	0.773726165314585\\
2.5	0.738733238808624\\
2.75	0.676422154699111\\
3	0.581782401923789\\
3.25	0.508656699895979\\
3.5	0.452950847763713\\
3.75	0.403555673363635\\
4	0.37051645208933\\
4.25	0.355739925003932\\
4.5	0.31729955522413\\
4.75	0.302261873636576\\
5	0.287233114907435\\
5.25	0.274703810029482\\
5.5	0.263741193410917\\
5.75	0.250795673118005\\
6	0.240202478632606\\
};
\addlegendentry{\tiny root-MUSIC}

\addplot [color=mycolor9, line width = 1pt, mark = square, mark options={solid, mycolor9}, mark repeat = 2]
  table[row sep=crcr]{%
0.5	39.7665894707892\\
0.75	36.8634010947122\\
1	38.5028963588649\\
1.25	39.9454455361646\\
1.5	41.4465548210063\\
1.75	42.746746516365\\
2	40.8662406429685\\
2.25	39.5011612026529\\
2.5	35.9609995081372\\
2.75	35.2819550687988\\
3	29.2312068916623\\
3.25	25.3919533605663\\
3.5	19.0553700920905\\
3.75	13.9719056583537\\
4	7.02409304824153\\
4.25	4.34335460885406\\
4.5	2.42748616674102\\
4.75	0.323917450339365\\
5	0.304282327751331\\
5.25	0.290992978612782\\
5.5	0.277128973499522\\
5.75	0.263005036912161\\
6	0.252152270772189\\};
\addlegendentry{\tiny PR-DML} 

\addplot [color=mycolor6, mark=asterisk, mark options={solid, mycolor6}, line width = 1pt, dashdotted, mark repeat = 2]
table[row sep=crcr]{%
0.5	38.4642296625556\\
0.75	29.4115252663978\\
1	27.3100578086433\\
1.25	29.559650551247\\
1.5	31.4954439765404\\
1.75	35.3270179100543\\
2	33.6398490605462\\
2.25	33.5194000498686\\
2.5	29.9204271581343\\
2.75	30.7363877009747\\
3	24.4476296213398\\
3.25	21.1040423829317\\
3.5	17.1877132192129\\
3.75	11.8628996114407\\
4	6.104134719801\\
4.25	4.2952185548856\\
4.5	0.341617761899709\\
4.75	0.320818604884548\\
5	0.302347510420771\\
5.25	0.291358690641562\\
5.5	0.277721239194931\\
5.75	0.26294670599158\\
6	0.251546065009569\\
};
\addlegendentry{\tiny PR-WSF}

\addplot [color=mycolor3, line width = 1pt, mark = triangle, mark options={solid, mycolor3}, mark repeat = 2, dashdotted]
table[row sep=crcr]{%
	0.5	43.2305836716323\\
	0.75	21.8047958366108\\
	1	5.27169103479491\\
	1.25	1.80607340998996\\
	1.5	1.48945277952776\\
	1.75	1.24952827027033\\
	2	1.12169664692669\\
	2.25	1.00320504993833\\
	2.5	0.890059300840773\\
	2.75	0.802222578182313\\
	3	0.704493095611143\\
	3.25	0.627521666205428\\
	3.5	0.567911274003948\\
	3.75	0.514751240643548\\
	4	0.474021603101878\\
	4.25	0.427276989425038\\
	4.5	0.39121531267866\\
	4.75	0.367242227930941\\
	5	0.3420961827913\\
	5.25	0.318945837207439\\
	5.5	0.300462163027473\\
	5.75	0.289599263996534\\
	6	0.27633642699374\\
};
\addlegendentry{\tiny PR-CCF}

\addplot [color=mycolor3, line width = 1pt,  mark repeat = 2]
table[row sep=crcr]{%
	0.5	43.4918417435825\\
	0.75	21.2510453372397\\
	1	5.22859937507921\\
	1.25	1.91220527526467\\
	1.5	1.60141140439275\\
	1.75	1.37185508346788\\
	2	1.22837261095509\\
	2.25	1.08993466046534\\
	2.5	0.954209091061417\\
	2.75	0.846209059604411\\
	3	0.736209667540244\\
	3.25	0.639121852124055\\
	3.5	0.571408550311251\\
	3.75	0.512065472431127\\
	4	0.470603605691653\\
	4.25	0.423496864258323\\
	4.5	0.383566498721894\\
	4.75	0.362619596458866\\
	5	0.334987890289819\\
	5.25	0.313166622665837\\
	5.5	0.294169605118217\\
	5.75	0.281520944711505\\
	6	0.269538331847118\\
};
\addlegendentry{\tiny PR-UCF}

\addplot [color=mycolor9, line width = 1pt, dashdotted]
table[row sep=crcr]{%
0.5	34.9196616361791\\
0.75	15.8317794957481\\
1	3.32921943544826\\
1.25	1.40132660109169\\
1.5	1.03585850785778\\
1.75	0.882209646499242\\
2	0.75762580779785\\
2.25	0.665183691885429\\
2.5	0.619109251593952\\
2.75	0.557680085494047\\
3	0.524947864428839\\
3.25	0.467382925250952\\
3.5	0.421713030010838\\
3.75	0.402336761070253\\
4	0.36041572222114\\
4.25	0.350044379268157\\
4.5	0.32964198920866\\
4.75	0.317992506613393\\
5	0.300994617419838\\
5.25	0.287901497328049\\
5.5	0.274273736207232\\
5.75	0.262734127970595\\
6	0.258330096168529\\
};
\addlegendentry{\tiny DML}

\addplot [color=black, line width = 1pt]
  table[row sep=crcr]{%
0.5	5.36077239986532\\
0.75	2.72855297209789\\
1	1.77276398834461\\
1.25	1.30587297156695\\
1.5	1.03504932732284\\
1.75	0.85955546617051\\
2	0.73684588450239\\
2.25	0.646219259646169\\
2.5	0.576499024677175\\
2.75	0.521151855373833\\
3	0.476112504498286\\
3.25	0.438721291828042\\
3.5	0.407165244338203\\
3.75	0.380165714699942\\
4	0.35679466997488\\
4.25	0.336361901434646\\
4.5	0.318343178020435\\
4.75	0.302332999334007\\
5	0.288012650264825\\
5.25	0.275128067721298\\
5.5	0.2634741706745\\
5.75	0.25288355053809\\
6	0.243218166642724\\
};
\addlegendentry{\tiny CRB}

\end{axis}
\end{tikzpicture}%
	\caption{Uncorrelated source signals, $\text{SNR} = 10 \text{ dB}$, number of snapshots $T = 100$}
	\label{fig:uncorrDelta}
\end{figure}
}
Figure~\ref{fig:uncorrDelta_lowNSnp} depicts a scenario where the number of snapshots $T = 8$ is smaller than the number of antennas $M=10$. In this case, the sample covariance matrix calculated in \eqref{eq:defCovMat} is singular, and therefore the PR-CCF is not applicable. In this case, we apply the diagonal loading technique with the loading factor $\gamma = 10^{-4}$ on the sample covariance matrix. The initialization of $\sigs{left}^2$ in Algorithm~\ref{alg:PR-UCF} is set at $ 10^{-6}$. To avoid outliers in RMSE caused by misdetection and to simulate the DOA tracking process \cite{HenHu2014799}, 1\% of the estimates with the largest error for all investigated algorithms are removed before calculating the RMSE. It can be observed that, even in the case of a very low number of snapshots, PR-UCF obtains a remarkable threshold behavior, outperforming other methods except for the brute-force DML. The RMSE of PR-UCF only slowly approaches the Cramer-Rao Bound as the SNR increases. In the high SNR regime, however, the RMSE of PR-UCF is very close to the Cramer-Rao Bound. The performance of PR-CCF is highly degraded due to the diagonal loading. Further research may be carried out regarding the optimal adaptive choice of the diagonal loading factor $\gamma$ to achieve an improved performance using a direction-dependent factor. However, this is beyond the scope of this paper, and therefore left for further research. Similar to the above-investigated scenarios, PR-DML and PR-WSF outperform MUSIC consistently.\\
{\setlength{\textfloatsep}{-20pt}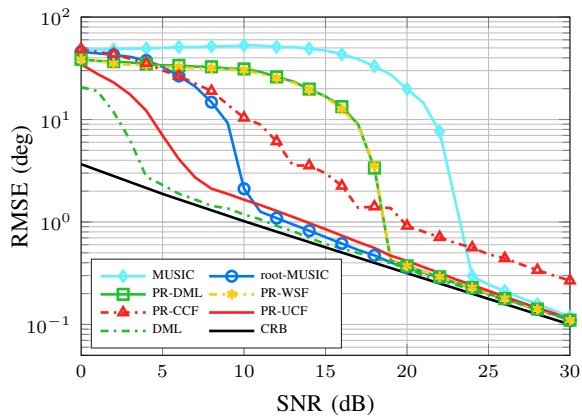
\begin{figure}[t]
	\centering
%
%
\definecolor{mycolor1}{RGB}{136, 0, 21}%
\definecolor{mycolor2}{RGB}{185,122,187}%
\definecolor{mycolor3}{RGB}{255,32,27}%
\definecolor{mycolor4}{RGB}{255,174,201}%
\definecolor{mycolor5}{RGB}{255,137,29}%
\definecolor{mycolor6}{RGB}{255,201,14}%
\definecolor{mycolor7}{RGB}{255,242,0}%
\definecolor{mycolor8}{RGB}{239,228,176}%
\definecolor{mycolor9}{RGB}{31,194,43}%
\definecolor{mycolor10}{RGB}{181,230,29}%
\definecolor{mycolor11}{RGB}{0,119,251}%
\definecolor{mycolor12}{RGB}{94,243,255}%

\definecolor{mycolor13}{RGB}{63,72,204}%
\definecolor{mycolor14}{RGB}{112,146,190}%
\definecolor{mycolor15}{RGB}{163,73,164}%
\definecolor{mycolor16}{RGB}{200,191,231}%
\begin{tikzpicture}

\begin{axis}[%
width=6.5cm,
height=4.5cm,
at={(0in,0in)},
scale only axis,
every tick label/.append style={font=\scriptsize},
xmin=0,
xmax=30,
xlabel={\small SNR (dB)},
ymode=log,
ymin=5e-2,
ymax=100,
ylabel={\small RMSE (deg)},
y label style={at={(axis description cs:-0.08,.5)},anchor=south},
yminorticks=true,
grid = both,
axis background/.style={fill=white},
legend columns = 2,
legend style={legend cell align=left, align=left, draw=white!15!black, row sep=-2pt, at={(0.02,0.02)}, anchor = south west}
]
\addplot [color=mycolor12, line width = 1pt, mark = diamond, mark options={solid, mycolor12}, mark repeat = 2]
  table[row sep=crcr]{%
0	48.3041348481575\\
1	48.3141276845115\\
2	48.4283201168679\\
3	49.2881152082484\\
4	49.3384814485015\\
5	49.9564386192526\\
6	51.0669384402862\\
7	50.8592881733876\\
8	51.8277451302927\\
9	51.834780464435\\
10	52.9324459329788\\
11	52.7502818919872\\
12	51.0468432689933\\
13	50.9409491866019\\
14	49.1918940447795\\
15	47.0988671457805\\
16	43.4571080774557\\
17	38.806011875168\\
18	33.109042333397\\
19	27.3864990063545\\
20	19.9094680647581\\
21	14.5266343096867\\
22	7.64974554492436\\
23	1.50971216882719\\
24	0.297270928846131\\
25	0.244127068243556\\
26	0.212057223261145\\
27	0.180748917793786\\
28	0.158075975404244\\
29	0.137376410553127\\
30	0.118976383550234\\
};
\addlegendentry{\tiny MUSIC}

\addplot [color=mycolor11, line width = 1pt, mark = o, mark options={solid, mycolor11}, mark repeat = 2]
  table[row sep=crcr]{%
0	45.7909231274017\\
1	44.2859703039961\\
2	42.8411245812261\\
3	40.8896317466758\\
4	37.6801594088904\\
5	32.7085512765782\\
6	26.5766559475297\\
7	20.9965075616246\\
8	14.7639143698177\\
9	9.24168134231787\\
10	2.10974404573436\\
11	1.25992954664741\\
12	1.09112995621693\\
13	0.94644694608737\\
14	0.821598055228232\\
15	0.712283239473622\\
16	0.615821339016373\\
17	0.533441543690288\\
18	0.472130349601665\\
19	0.403351132181867\\
20	0.359426010731399\\
21	0.321873553670657\\
22	0.285637101454974\\
23	0.252813422983993\\
24	0.225973161376473\\
25	0.199106210724148\\
26	0.177350350948096\\
27	0.15697637634798\\
28	0.139927236176533\\
29	0.12505822419784\\
30	0.108732373586638\\
};
\addlegendentry{\tiny root-MUSIC}

\addplot [color=mycolor9, line width = 1pt, mark = square, mark options={solid, mycolor9}, mark repeat = 2]
  table[row sep=crcr]{%
0	38.6966793000833\\
1	37.3288860857803\\
2	36.7021373274727\\
3	36.0256340848165\\
4	34.7511063756565\\
5	33.7727099898142\\
6	33.7269310614934\\
7	32.8113054577556\\
8	32.720943101247\\
9	31.4788900380368\\
10	31.1441674950198\\
11	29.1677371228856\\
12	26.0252996239328\\
13	23.5437884774534\\
14	19.8255262782887\\
15	16.7998529350372\\
16	13.3505585782989\\
17	8.95160298396854\\
18	3.35931981778029\\
19	0.423136456644283\\
20	0.372924301268015\\
21	0.328352253788325\\
22	0.291952989536152\\
23	0.255065535523074\\
24	0.229544187217395\\
25	0.201644732509668\\
26	0.17899622281227\\
27	0.1586846302952\\
28	0.141373054279808\\
29	0.126629940151214\\
30	0.110177542112973\\
};
\addlegendentry{\tiny PR-DML}

\addplot [color=mycolor6, mark=asterisk, mark options={solid, mycolor6}, line width = 1pt, dashdotted, mark repeat = 2]
table[row sep=crcr]{%
0	38.8253597877983\\
1	37.3275468335122\\
2	35.2063790142247\\
3	34.2333531082826\\
4	33.9011648437449\\
5	32.7673560049921\\
6	32.514135713061\\
7	31.5421133610867\\
8	31.7266679042919\\
9	30.7299898164538\\
10	30.3253048901727\\
11	27.7751772470514\\
12	25.381117785123\\
13	22.9516107439223\\
14	19.5554861837094\\
15	16.5249432596491\\
16	12.6719687489755\\
17	8.67650605224644\\
18	3.53408306992511\\
19	0.422618319570518\\
20	0.372624210960848\\
21	0.32827890304943\\
22	0.291907190141203\\
23	0.254900521241051\\
24	0.229471825909541\\
25	0.201628503461029\\
26	0.178992840895141\\
27	0.158683358701562\\
28	0.141348105381678\\
29	0.126623981802706\\
30	0.110151819577414\\
};
\addlegendentry{\tiny PR-WSF}

\addplot [color=mycolor3, line width = 1pt, mark = triangle, mark options={solid, mycolor3}, mark repeat = 2, dashdotted]
table[row sep=crcr]{%
	0	48.7839443878256\\
	1	45.690063198253\\
	2	43.5913422243364\\
	3	39.1207030990584\\
	4	35.5126535879583\\
	5	30.148220836237\\
	6	26.5281667285086\\
	7	22.4592063700655\\
	8	18.9944571343034\\
	9	13.9600599229917\\
	10	10.3984899079302\\
	11	8.92175860417232\\
	12	6.10895231636559\\
	13	3.56358629378359\\
	14	3.55166175915099\\
	15	2.99102603041846\\
	16	2.25204678620482\\
	17	1.37398404728184\\
	18	1.41316600175865\\
	19	1.36352763939867\\
	20	0.916543180504201\\
	21	0.806193674683906\\
	22	0.710990698429846\\
	23	0.620887776343348\\
	24	0.563366355742365\\
	25	0.483160244047453\\
	26	0.443007036827389\\
	27	0.385292220919303\\
	28	0.339394898696782\\
	29	0.300411872301492\\
	30	0.267653673464504\\
};
\addlegendentry{\tiny PR-CCF}

\addplot [color=mycolor3, line width = 1pt,  mark repeat = 2]
table[row sep=crcr]{%
0	34.6688779541892\\
1	27.9135815710907\\
2	22.9863430093786\\
3	17.6000989055128\\
4	12.1969517895253\\
5	6.99368624814905\\
6	4.11583517227714\\
7	2.71846416278537\\
8	2.11385047095731\\
9	1.87997249348486\\
10	1.64975060584369\\
11	1.46194028620829\\
12	1.28101221353882\\
13	1.12038496589364\\
14	0.976121814529123\\
15	0.849566163398451\\
16	0.736921432398646\\
17	0.63746256918679\\
18	0.558094765224452\\
19	0.468214688412315\\
20	0.417449536432189\\
21	0.36307127528361\\
22	0.318661108500954\\
23	0.274999887297396\\
24	0.243034593074387\\
25	0.212917040227095\\
26	0.186587180333481\\
27	0.164262205004185\\
28	0.144836387760192\\
29	0.129007870373658\\
30	0.111663475078511\\
};
\addlegendentry{\tiny PR-UCF}

\addplot [color=mycolor9, line width = 1pt, dashdotted]
  table[row sep=crcr]{%
0	20.7212084658379\\
1	18.9674011109538\\
2	11.8249977901592\\
3	6.3226285330044\\
4	2.76432749274293\\
5	2.29927504713555\\
6	1.88683791236216\\
7	1.65684836279088\\
8	1.43773924325521\\
9	1.35684581995017\\
10	1.18196737805742\\
11	1.06033447800994\\
12	0.916005219999474\\
13	0.814065028966328\\
14	0.700368095132032\\
15	0.610083457840121\\
16	0.550978680773068\\
17	0.497679865473547\\
18	0.455125240631846\\
19	0.392972006358363\\
20	0.356839664108791\\
21	0.299649581358994\\
22	0.268694930288705\\
23	0.243503851158016\\
24	0.2172039891172\\
25	0.192837556053498\\
26	0.173780712036531\\
27	0.160081262317087\\
28	0.136853877436562\\
29	0.123139155665173\\
30	0.113470308669766\\
};
\addlegendentry{\tiny DML}

\addplot [color=black, line width = 1pt]
  table[row sep=crcr]{%
0	3.6672884239183\\
5	1.87948644754234\\
10	1.01827849034884\\
15	0.565275273400639\\
20	0.316541387177053\\
25	0.177764911692747\\
30	0.099921884799054\\
35	0.056182611290586\\
40	0.0315924533715512\\
};
\addlegendentry{\tiny CRB}

\end{axis}
\end{tikzpicture}%
	\caption{Uncorrelated source signals, number of snapshots $T = 8$}
	\label{fig:uncorrDelta_lowNSnp}
\end{figure}
}
In Figure~\ref{fig:runTime}, the execution time of the DOA estimation algorithms with respect to the number of antennas $M$ are depicted. We do not include the brute-force DML due to the high execution time. The angle-of-view is partitioned uniformly into $N_G = 1800$ directions. The term \textit{Generic} in Figure~\ref{fig:runTime} refers to the naive implementation using the MATLAB command \texttt{eig} for the eigenvalue decomposition. The rooting process applied to root-MUSIC relies on determining the eigenvalues of the companion matrix associated with the polynomial, and therefore the execution time increases drastically with respect to the number of antennas $M$. All partial relaxation methods, except for the PR-UCF estimator, follow similar trends as MUSIC, where the execution time is in the same order of magnitude. The execution time of the PR-UCF estimator is approximately ten times larger than other partial relaxation methods. Nevertheless, the PR-UCF estimator based on the efficient eigenvalue decomposition introduced in Section~\ref{sec:computationalAspect} requires less execution time than the direct implementation with the MATLAB command. Generally, thanks to the quadratic convergence behavior of Algorithm~\ref{alg:RootingSecularFunction}, the execution time is reduced by a factor of 20 to 1000 in comparison with the direct implementation using the generic command \texttt{eig}. PR-WSF even exhibits almost identical execution time behavior as MUSIC, indicating the possibility of applying the partial relaxation methods in practical cases.
\begin{figure}[t]
	\centering
%
%
\definecolor{mycolor1}{RGB}{136, 0, 21}%
\definecolor{mycolor2}{RGB}{185,122,187}%
\definecolor{mycolor3}{RGB}{255,32,27}%
\definecolor{mycolor4}{RGB}{255,174,201}%
\definecolor{mycolor5}{RGB}{255,137,29}%
\definecolor{mycolor6}{RGB}{255,201,14}%
\definecolor{mycolor7}{RGB}{255,242,0}%
\definecolor{mycolor8}{RGB}{239,228,176}%
\definecolor{mycolor9}{RGB}{31,194,43}%
\definecolor{mycolor10}{RGB}{181,230,29}%
\definecolor{mycolor11}{RGB}{0,119,251}%
\definecolor{mycolor12}{RGB}{94,243,255}%

\definecolor{mycolor13}{RGB}{63,72,204}%
\definecolor{mycolor14}{RGB}{112,146,190}%
\definecolor{mycolor15}{RGB}{163,73,164}%
\definecolor{mycolor16}{RGB}{200,191,231}%
\begin{tikzpicture}

\begin{axis}[%
width=6.5cm,
height=4.5cm,
at={(0in,0in)},
scale only axis,
every tick label/.append style={font=\scriptsize},
xmin=5,
xmax=50,    
ymode=log,
ymin=0.000005,
ymax=50,
xlabel={\small Number of Antennas $M$ },
ylabel={\small Execution time (s)},
y label style={at={(axis description cs:-0.09,.5)},anchor=south},
grid = both,
yminorticks=true,
axis background/.style={fill=white},
legend columns=2, 
legend style={legend cell align=left, align=left, draw=white!15!black, row sep=-2pt, at={(0.98,0.02)}, anchor = south east}
]
\addplot [color=mycolor12, line width = 1pt, mark = diamond, mark options={solid, mycolor12}]
  table[row sep=crcr]{%
5	0.000904067067067067\\
10	0.00089048948948949\\
15	0.00105174674674675\\
20	0.00113523023023023\\
25	0.00128054154154154\\
30	0.00142466666666667\\
35	0.00158354154154154\\
40	0.00171751251251251\\
45	0.00188324824824825\\
50	0.00205294594594595\\
};
\addlegendentry{\tiny MUSIC}

\addplot [color=mycolor11, line width = 1pt, mark = o, mark options={solid, mycolor11}]
  table[row sep=crcr]{%
5	0.000149901901901902\\
10	0.000326050050050051\\
15	0.00062274074074074\\
20	0.00123996896896897\\
25	0.00205020920920921\\
30	0.00194845545545545\\
35	0.00277786286286286\\
40	0.00632194294294294\\
45	0.00770872272272271\\
50	0.0079693973973974\\
};
\addlegendentry{\tiny root-MUSIC}

\addplot [color=mycolor9, line width = 1pt, mark = square, mark options={solid, mycolor9}]
  table[row sep=crcr]{%
5	0.00142514141414141\\
10	0.00158118181818182\\
15	0.00197354545454545\\
20	0.00231682828282828\\
25	0.00271547474747475\\
30	0.00315049494949495\\
35	0.00356409090909091\\
40	0.0038810101010101\\
45	0.00438780808080808\\
50	0.00478877777777778\\
};
\addlegendentry{\tiny PR-DML} 

\addplot [color=mycolor9, line width = 1pt, mark = square, mark options={solid, mycolor9}, dashdotted]
table[row sep=crcr]{%
5	0.0341984343434343\\
10	0.0670773333333333\\
15	0.163646161616162\\
20	0.257155080808081\\
25	0.378302828282828\\
30	0.531098737373737\\
35	0.847944252525252\\
40	1.04346464646465\\
45	1.31501480808081\\
50	1.62496153535354\\
};
\addlegendentry{\tiny Generic PR-DML} 

\addplot [color=mycolor6, mark=asterisk, mark options={solid, mycolor6}, line width = 1pt]
table[row sep=crcr]{%
5	0.00115768686868687\\
10	0.00109747474747475\\
15	0.00123238383838384\\
20	0.0013880303030303\\
25	0.00159462626262626\\
30	0.00178194949494949\\
35	0.00197867676767677\\
40	0.00206340404040404\\
45	0.00226557575757576\\
50	0.00244640404040404\\
};
\addlegendentry{\tiny PR-WSF}

\addplot [color=mycolor6, mark=asterisk, mark options={solid, mycolor6}, line width = 1pt, dashdotted]
table[row sep=crcr]{%
5	0.0324110909090909\\
10	0.0716349090909091\\
15	0.228684505050505\\
20	0.390904090909091\\
25	0.595977838383838\\
30	0.821700353535353\\
35	1.22924973737374\\
40	1.67382903030303\\
45	2.09726166666667\\
50	2.60814714141414\\
};
\addlegendentry{\tiny Generic PR-WSF}

\addplot [color=mycolor3, line width = 1pt, mark = triangle, mark options={solid, mycolor3}]
  table[row sep=crcr]{%
5	0.0013820101010101\\
10	0.00162925252525252\\
15	0.00203233333333333\\
20	0.0023960202020202\\
25	0.00292831313131313\\
30	0.00345309090909091\\
35	0.00401219191919192\\
40	0.00414719191919192\\
45	0.00469017171717172\\
50	0.00524051515151515\\
};
\addlegendentry{\tiny PR-CCF}

\addplot [color=mycolor3, line width = 1pt, mark = triangle, mark options={solid, mycolor3}, dashdotted]
table[row sep=crcr]{%
5	0.0246795252525253\\
10	0.0545189797979798\\
15	0.142212252525252\\
20	0.219713161616162\\
25	0.324736343434343\\
30	0.457240525252525\\
35	0.623050727272727\\
40	0.820083343434343\\
45	1.05898157575758\\
50	1.33167343434343\\
};
\addlegendentry{\tiny Generic PR-CCF}

\addplot [color=mycolor3, line width = 1pt]
table[row sep=crcr]{%
5	0.0111443853853854\\
10	0.0139562902902903\\
15	0.0170504314314314\\
20	0.0199981831831832\\
25	0.0227646236236236\\
30	0.0261032892892893\\
35	0.0270102252252252\\
40	0.0286304974974975\\
45	0.0326097697697698\\
50	0.0363567037037038\\
};
\addlegendentry{\tiny PR-UCF}

\addplot [color=mycolor3, line width = 1pt, dashdotted]
table[row sep=crcr]{%
5	0.628210636363636\\
10	1.44332188888889\\
15	3.76705633333333\\
20	6.03818023232323\\
25	8.95405080808081\\
30	12.4179207373737\\
35	17.0191143030303\\
40	23.1836166060606\\
45	30.2411949494949\\
50	37.9566288181818\\
};
\addlegendentry{\tiny Generic PR-UCF}

\end{axis}
\end{tikzpicture}%
	\caption{$\text{SNR} = 10 \text{ dB}$, number of snapshots $T = 100$}
	\label{fig:runTime}
\end{figure}
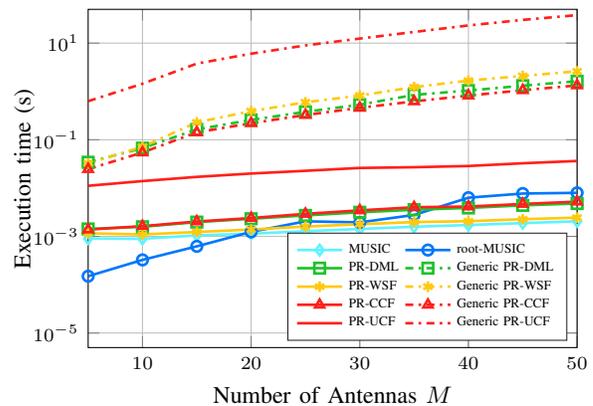
\section{Conclusions and Outlook}\label{sec:Conclusion}
In this paper, new DOA estimators using the partial relaxation approach are introduced. Instead of enforcing the full structure on the steering matrix when formulating the DOA estimation problem, in the partial relaxation approach, only the structure in the steering vector of one source of interest is preserved, while the structure of the remaining interfering sources is relaxed. The null-spectra of the partial relaxation methods are efficiently calculated by applying known results regarding the rank-one modification of a Hermitian matrix. Simulation results show that, in the proposed framework, even though no particular structure of the sensor array, e.g., Vandermonde structure from a uniform linear array, is required, the proposed methods based on the covariance fitting problems exhibit comparable threshold performance as DML, and superior to spectral MUSIC in difficult scenarios. As a result, the estimates obtained from the proposed covariance fitting problems can be employed as initializations for solving the maximum likelihood problems. In comparison with the unconstrained covariance fitting estimator, the inner approximation in the constrained version helps to reduce the computational complexity without necessarily sacrificing the error performance. One weakness of both covariance fitting variants is the slight deviation from the Cramer-Rao Bound in the asymptotic region. Although the performance in the non-asymptotic region is not as remarkable as the covariance fitting variants, the proposed estimator based on the Weighted Subspace Fitting problem is still favorable in certain circumstances due to the low computational complexity and the excellent asymptotic behavior.

For future work, the theoretical error behavior and consistency of methods in the family of the partial relaxation approach is an interesting open problem and requires further investigation.

\appendices
\section{Proof of \eqref{eq:mainPRDMLresult}}\label{appsec:PRDMLproof}
In order to prove the identity in \eqref{eq:mainPRDMLresult}, we first introduce two important prepositions:
\vspace*{14pt}
\begin{Proposition}\label{prop1}
	Let $\vec{B}\in \complexset^{M\times(N-1)}$ be a non-zero matrix and $\vec{a}\in \complexset^{M\times 1}$ be a non-zero vector such that $\oproj{\vec{a}}{\vec{B}}$ is a non-zero matrix. Then there always exists a matrix ${\vec{Z}\in\complexset^{M\times N'}}$ with $1\leq N'\leq N-1$ such that the following conditions are satisfied:\vspace*{-7pt}
	\begin{subequations}\label{eq:orthodecomposition}
	\begin{align}
	\label{eq:firstcondition}
	\proj{\oproj{\vec{a}}{\vec{B}}}{} &= \vec{Z}\vec{Z}^H\\
	\label{eq:secondcondition}
	\vec{Z}^H\vec{Z} &= \vec{I}_{N'}\\
	\label{eq:thirdcondition}
	\vec{Z}^H\vec{a} &= \vec{0}.
	\end{align}
	\end{subequations}
\end{Proposition}
\begin{IEEEproof}[Proof of Proposition~\ref{prop1}]
	First expand $\proj{\oproj{\vec{a}}{\vec{B}}}{}$ in the following manner:\vspace*{-7pt}
	\begin{equation}
	\begin{aligned}
	\proj{\oproj{\vec{a}}{\vec{B}}}{} &= \oproj{\vec{a}}{\vec{B}}\mathopen{}\left(\left(\oproj{\vec{a}}{\vec{B}}\right)^H\left(\oproj{\vec{a}}{\vec{B}}\right)\right)^{-1}\mathclose{}\left(\oproj{\vec{a}}{\vec{B}}\right)^H \\
	&= \oproj{\vec{a}}{\vec{B}}\left(\vec{B}^H\oproj{\vec{a}}{\vec{B}}\right)^{-1}\vec{B}^H\oproj{\vec{a}}{}.
	\end{aligned}
	\end{equation}
	Therefore, the rank of the matrix $\proj{\oproj{\vec{a}}{\vec{B}}}{}$ is bounded by:\vspace*{-4pt}
\begin{equation}
\label{eq:rankBound}
\begin{aligned}
1&\leq N' = \text{rank}\left(\proj{\oproj{\vec{a}}{\vec{B}}}{}\right)\\
&\leq\min\mathopen{}\left\lbrace\text{rank}\left(\oproj{\vec{a}}{}\right), \text{rank}\left(\vec{B}\right),  \text{rank}\left(\left(\vec{B}^H\oproj{\vec{a}}{\vec{B}}\right)\right)\right\rbrace\mathclose{}\\
&\leq N - 1
\end{aligned}
\end{equation}
	Furthermore, since $\proj{\oproj{\vec{a}}{\vec{B}}}{}$ contains only the eigenvalues $0$ and $1$, taking the eigenvalue decomposition of $\proj{\oproj{\vec{a}}{\vec{B}}}{}$ leads to:
	\begin{equation}
	\label{eq:findZ}
	\proj{\oproj{\vec{a}}{\vec{B}}}{} = \vec{Z}\vec{Z}^H, 
	\end{equation}
	and the number of columns of $\vec{Z}$ is equal to $N'$. Clearly the matrix $\vec{Z}\in\complexset^{M\times N'}$ satisfies $\vec{Z}^H\vec{Z} = \vec{I}_{N'}$. Therefore, the conditions in \eqref{eq:firstcondition} and \eqref{eq:secondcondition} are satisfied if $\vec{Z}$ is chosen as in \eqref{eq:findZ}. Finally, we observe that:
	\begin{equation}\label{eq:normZha}
	\begin{aligned}
	\vec{a}^H\vec{Z}\vec{Z}^H\vec{a} &= \vec{a}^H\proj{\oproj{\vec{a}}{\vec{B}}}{\vec{a}}\\ &= \vec{a}^H\oproj{\vec{a}}{\vec{B}}\mathopen{}\left(\vec{B}^H\oproj{\vec{a}}{\vec{B}}\right)^{-1}\mathclose{}\vec{B}^H\oproj{\vec{a}}{\vec{a}} = 0
	\end{aligned}
	\end{equation}
	The identity in \eqref{eq:thirdcondition} follows immediately from \eqref{eq:normZha}.
\end{IEEEproof}
\vspace*{7pt}
\begin{Proposition}\label{prop2}
	Let $\vec{a}\in\complexset^{M\times 1}$ be a non-zero vector and $\hat{\vec{R}}\in\complexset^{M\times M}$ be a non-zero Hermitian positive semidefinite matrix. Then the eigenvectors which correspond to non-zeros eigenvalues of $\oproj{\vec{a}}{\hat{\vec{R}}}\oproj{\vec{a}}{}$ are orthogonal to $\vec{a}$.
\end{Proposition}
\begin{IEEEproof}[Proof of Proposition~\ref{prop2}]
	Let $\vec{x}$ be an eigenvector corresponding to an eigenvalue $\lambda\neq 0$ of $\oproj{\vec{a}}{\hat{\vec{R}}}\oproj{\vec{a}}{}$, then by definition:
	\begin{equation}
	\label{eq:defEig}
	\oproj{\vec{a}}{\hat{\vec{R}}}\oproj{\vec{a}}{} \vec{x} = \lambda\vec{x}.
	\end{equation}
	Taking the conjugate transpose of \eqref{eq:defEig} and multiplying with $\vec{a}$ on the right, we obtain:
	\begin{equation}\label{eq:proofOrtho}
	0 = \vec{x}^H\oproj{\vec{a}}{\hat{\vec{R}}}\oproj{\vec{a}}{\vec{a}} = \left(\oproj{\vec{a}}{\hat{\vec{R}}}\oproj{\vec{a}}{} \vec{x}\right)^*\vec{a} =  \lambda^*\vec{x}^H\vec{a}.
	\end{equation}
	From \eqref{eq:proofOrtho} and the assumption that $\lambda$ is non-zero, we can conclude that $\vec{x}$ is orthogonal to $\vec{a}$.
\end{IEEEproof}
Now we return to the main proof of \eqref{eq:mainPRDMLresult}. In the case that $\oproj{\vec{a}}{\vec{B}} = \vec{0}$, then by convention in Section~\ref{subsec:derivationPRDML}, we obtain that $\tr{\proj{\oproj{\vec{a}}{\vec{B}}}{\hat{\vec{R}}}} = 0$. On the other hand, if the matrix $\oproj{\vec{a}}{\vec{B}}$ is a non-zero matrix, applying the decomposition \eqref{eq:orthodecomposition} in Proposition~\ref{prop1} and noting that $\vec{Z}^H\vec{a} = 0$, the objective function in \eqref{eq:mainPRDMLresult} is rewritten as follows:
\begin{equation}
\label{eq:reform1}
	\begin{aligned}
	&\tr{\proj{\oproj{\vec{a}}{\vec{B}}}{\hat{\vec{R}}}} = \tr{\vec{Z}\vec{Z^H}\hat{\vec{R}}} = \tr{\vec{Z}^H\hat{\vec{R}}\vec{Z}}\\
	=\text{ } &\tr{\vec{Z}^H\left(\proj{\vec{a}}{} + \oproj{\vec{a}}{}\right)\hat{\vec{R}}\left(\proj{\vec{a}}{} + \oproj{\vec{a}}{}\right)\vec{Z}}\\
	=\text{ } &\tr{\vec{Z}^H\oproj{\vec{a}}{\hat{\vec{R}}}\oproj{\vec{a}}{\vec{Z}}}.
	\end{aligned}
\end{equation}
	Therefore, the optimization problem in \eqref{eq:mainPRDMLresult} is reformulated as:
	\vspace*{-10pt}
	\begin{subequations}
		\label{eq:reformulatedRLS}
	\begin{align}
	&\underset{\vec{Z} \in \complexset^{M\times N'}}{\text{maximize }} \tr{\vec{Z}^H\oproj{\vec{a}}{\hat{\vec{R}}}\oproj{\vec{a}}{\vec{Z}}}\\
	&\text{subject to }\vec{Z}^H\vec{Z} = \vec{I}_{N'}\\
	\label{eq:thirdConstraint}&\text{\color{white}subject to } \vec{Z}^H\vec{a} = \vec{0}.		\vspace*{-10pt}
	\end{align}
		\end{subequations}
Dropping the constraint $\vec{Z}^H\vec{a} = 0$ in \eqref{eq:thirdConstraint}, we obtain the relaxed optimization problem: 
	\begin{subequations}
	\label{eq:relaxedProb}
	\begin{align}
&\underset{\vec{Z} \in \complexset^{M\times N'}}{\text{maximize }} \tr{\vec{Z}^H\oproj{\vec{a}}{\hat{\vec{R}}}\oproj{\vec{a}}{\vec{Z}}}\\
&\text{subject to }\vec{Z}^H\vec{Z} = \vec{I}_{N'}
	\vspace*{-10pt}
	\end{align}
	\end{subequations}
From the Ky-Fan inequality in \cite{KyFanInequality}, the optimization in the relaxed problem in \eqref{eq:relaxedProb} admits a maximizer $\hat{\vec{Z}}$ whose columns form an orthonormal basis of the eigenspace associated with the $N'$-largest eigenvalues of $\oproj{\vec{a}}{\hat{\vec{R}}}\oproj{\vec{a}}{}$. However, Proposition~\ref{prop2} implies that any maximizer $\hat{\vec{Z}}$ of \eqref{eq:relaxedProb} also satisfies \eqref{eq:thirdConstraint}, i.e., $\hat{\vec{Z}}^H\vec{a} = 0$. Therefore, any maximizer $\hat{\vec{Z}}$ of the optimization problem in \eqref{eq:relaxedProb} is also a maximizer of \eqref{eq:reformulatedRLS}. As a consequence, we obtain the following result:
	\begin{equation}\label{eq:relaxedreformulatedRLS}
		\begin{aligned}
		\sum\limits_{k=1}^{N'}\lambda_k\left(\oproj{\vec{a}}{\hat{\vec{R}}}\oproj{\vec{a}}{}\right) = & \text{ max } \tr{\vec{Z}^H\oproj{\vec{a}}{\hat{\vec{R}}}\oproj{\vec{a}}{\vec{Z}}}\\[-1em]
		&\text{ subject to } \vec{Z} \in \complexset^{M\times N'}\\
		&\text{ \color{white}subject to } \vec{Z}^H\vec{Z} = \vec{I}_{N'},\\
		&\text{ \color{white}subject to } \vec{Z}^H\vec{a} = \vec{0}.
		\end{aligned}
	\end{equation}
	Combining \eqref{eq:rankBound}, \eqref{eq:reform1} and \eqref{eq:relaxedreformulatedRLS}, the following identity is obtained:
	\begin{equation}
	\label{eq:endRes}
	\begin{split}
	\underset{\vec{B}\in\complexset^{M\times(N-1)}}{\text{max }}\tr{\proj{\oproj{\vec{a}}{\vec{B}}}{\hat{\vec{R}}}} &= \sum\limits_{k = 1}^{N-1} \lambda_k\left(\oproj{a}{\hat{\vec{R}}\oproj{a}{}}\right)\\ &= \sum\limits_{k = 1}^{N-1} \lambda_k\left(\oproj{a}{\hat{\vec{R}}}\right).
	\end{split}
	\end{equation}
	The optimum in \eqref{eq:endRes} is achieved if we choose one matrix $\vec{B}\in\complexset^{M\times(N-1)}$ such that $\proj{\oproj{\vec{a}}{\vec{B}}}{} = \vec{Z}\vec{Z}^H$ and the columns of $\vec{Z}$ form an orthonormal basis of the eigenspace associated with $(N-1)$-principal eigenvalues of $\oproj{\vec{a}}{\hat{\vec{R}}}\oproj{\vec{a}}{}$.
	\section{Equivalence of MUSIC and PR-WSF with $\vec{W} = \vec{I}_N$}\label{appsec:PRSF}
	Considering the expression in \eqref{eq:PR-WSFPseudospectrum} for the steering vector ${\vec{a}=\vec{a}(\vartheta)}$, we note that the rank of  $\oproj{\vec{a}}{\hat{\vec{U}}_{\text{s}}\hat{\vec{U}}_{\text{s}}^H}$ is at most $N$. Hence, ${\lambda_k\left(\oproj{\vec{a}}{\hat{\vec{U}}_{\text{s}}\hat{\vec{U}}_{\text{s}}^H}\right) = 0}$ for ${k=N+1,\ldots,M}$. Therefore, when calculating the null-spectrum in \eqref{eq:PR-WSFPseudospectrum}, only $\lambda_N\left(\oproj{\vec{a}}{\hat{\vec{U}}_{\text{s}}\hat{\vec{U}}_{\text{s}}^H}\right)$ is considered. The expression for $\lambda_N\left(\oproj{\vec{a}}{\hat{\vec{U}}_{\text{s}}\hat{\vec{U}}_{\text{s}}^H}\right)$ can be further rewritten as follows:
	\begin{equation}
	\begin{aligned}
	\hspace*{-5pt}\lambda_N\mathopen{}\left(\oproj{\vec{a}}{\hat{\vec{U}}_{\text{s}}\hat{\vec{U}}_{\text{s}}^H}\right)\mathclose{}&=\lambda_N\mathopen{}\left(\hat{\vec{U}}_{\text{s}}^H\mathopen{}\left(\vec{I}_M - \dfrac{1}{\norm{\vec{a}}_2^2}\vec{a}\vec{a}^H\right)\mathclose{}\hat{\vec{U}}_{\text{s}}\right)\mathclose{}\\
	&=\lambda_N\left(\vec{I}_M - \dfrac{1}{\norm{\vec{a}}^2}\hat{\vec{U}}_{\text{s}}^H\vec{a}\vec{a}^H\hat{\vec{U}}_{\text{s}}\right)\\
	\label{eq:reformEig1}&= 1 + \lambda_N\left(-\dfrac{1}{\norm{\vec{a}}^2}\hat{\vec{U}}_{\text{s}}^H\vec{a}\vec{a}^H\hat{\vec{U}}_{\text{s}}\right).
	\end{aligned}
	\end{equation}
	Since ${-\dfrac{1}{\norm{\vec{a}}^2}\hat{\vec{U}}_{\text{s}}^H\vec{a}\vec{a}^H\hat{\vec{U}}_{\text{s}}}$ is a negative semidefinite rank-one matrix of size $N\times N$, it can be easily shown that:
	\begin{equation}\label{eq:smallestEig}
	\lambda_N\mathopen{}\left(-\dfrac{1}{\norm{\vec{a}}^2}\hat{\vec{U}}_{\text{s}}^H\vec{a}\vec{a}^H\hat{\vec{U}}_{\text{s}}\right)\mathclose{} = -\dfrac{1}{\norm{\vec{a}}_2^2}\vec{a}^H\hat{\vec{U}}_{\text{s}}\hat{\vec{U}}_{\text{s}}^H\vec{a}.
	\end{equation}
	Substituting \eqref{eq:smallestEig} into \eqref{eq:reformEig1} and using the orthogonality property between the signal and the noise subspace, we obtain:
	\begin{equation}
	\begin{aligned}
	\lambda_N\mathopen{}\left(\oproj{\vec{a}}{\hat{\vec{U}}_{\text{s}}\hat{\vec{U}}_{\text{s}}^H}\right)\mathclose{}&= 1 -\dfrac{1}{\norm{\vec{a}}_2^2}\vec{a}^H\mathopen{}\left(\vec{I}_M - \hat{\vec{U}}_{\text{n}}\hat{\vec{U}}_{\text{n}}^H\right)\mathclose{}\vec{a}\\
	\label{eq:equivPRSF}&= \dfrac{\vec{a}^H\hat{\vec{U}}_{\text{n}}\hat{\vec{U}}_{\text{n}}^H\vec{a}}{\vec{a}^H\vec{a}}.
	\end{aligned}
	\end{equation}
	The expression in \eqref{eq:equivPRSF} is identical to the null-spectrum of MUSIC.  Therefore, with $\vec{W} = \vec{I}_N$, the expression in \eqref{eq:SWSF} is another equivalent formulation of the MUSIC estimator.
	\section{Proof of \eqref{eq:derivUCF}}\label{appsec:derivEig}
	First, by taking the eigenvalue decomposition as in \eqref{eq:eigdecsample} and substituting $\vec{z} = \hat{\vec{U}}^H\vec{a}$, the inner objective function of the PR-UCF in \eqref{eq:reformulatedUCP} is rewritten as:
	\begin{equation}
	\label{eq:defObj}
	g(\sigma_{\text{s}}^2) = \sum\limits_{k = N}^{M}\lambda_k^2\left(\hat{\vec{\Lambda}} - \sigma_{\text{s}}^2\vec{z}\vec{z}^H\right).
	\end{equation}
	In the following steps, we calculate the derivative $\dfrac{d\lambda_k\left(\hat{\vec{\Lambda}} - \sigma_{\text{s}}^2\vec{z}\vec{z}^H\right)}{d\sigma_{\text{s}}^2}$. Applying the results from \cite{magnus1985differentiating} leads to the following expression:
	\begin{subequations}
		\vspace*{-7pt}
	\begin{align}
	\dfrac{d\lambda_k\left(\hat{\vec{\Lambda}} - \sigma_{\text{s}}^2\vec{z}\vec{z}^H\right)}{d\sigma_{\text{s}}^2} &= \dfrac{\bar{\vec{u}}_k^H\dfrac{d\left(\hat{\vec{\Lambda}} - \sigma_{\text{s}}^2\vec{z}\vec{z}^H\right)}{d\sigma_{\text{s}}^2}\bar{\vec{u}}_k}{\bar{\vec{u}}_k^H\bar{\vec{u}}_k}\\
	\label{eq:defDeriv}&= - \dfrac{\bar{\vec{u}}_k^H\vec{z}\vec{z}^H\bar{\vec{u}}_k}{\bar{\vec{u}}_k^H\bar{\vec{u}}_k},
	\end{align}
	\end{subequations}
	where $\bar{\vec{u}}_k$ is an eigenvector corresponding to the eigenvalue $\lambda_k\left(\hat{\vec{\Lambda}} - \sigma_{s}^2\vec{z}\vec{z}^H\right)$. Interestingly, the expression on the numerator of \eqref{eq:defDeriv} can be shown to be independent of the eigenvectors. In fact, by using the shorthand notation ${\bar{\lambda}_k(\sigma_{\text{s}}^2) = \lambda_k\left(\hat{\vec{\Lambda}} - \sigma_{\text{s}}^2\vec{z}\vec{z}^H\right)}$ as in \eqref{eq:shorthandLambda}, and applying Property 1 and Property 3 from Theorem~\ref{th:interlacing} to the matrix ${\hat{\vec{\Lambda}} - \sigma_{\text{s}}^2\vec{z}\vec{z}^H}$, we obtain:
	\begin{align}
	\label{eq:th1r1}0 &= 1 - \sigma_{\text{s}}^2\vec{z}^H\left(\hat{\vec{\Lambda}} - \bar{\lambda}\left(\sigma_{\text{s}}^2\right)\vec{I}_M\right)^{-1}\vec{z}\\
	\label{eq:th1r3}\bar{\vec{u}}_k &= \left(\hat{\vec{\Lambda}} - \bar{\lambda}_k\left(\sigma_{\text{s}}^2\right)\vec{I}_M\right)^{-1}\vec{z}.
	\end{align}
	Substituting \eqref{eq:th1r1} and \eqref{eq:th1r3} into \eqref{eq:defDeriv}, the derivative of $\lambda_k\left(\hat{\vec{\Lambda}} - \sigma_{\text{s}}^2\vec{z}\vec{z}^H\right)$ with respect to $\sigma_{\text{s}}^2$ is given by:
	\begin{gather}\label{eq:derivStep}
	\begin{aligned}
		&\dfrac{d\lambda_k\left(\hat{\vec{\Lambda}} - \sigma_{\text{s}}^2\vec{z}\vec{z}^H\right)}{d\sigma_{\text{s}}^2} = - \dfrac{\bar{\vec{u}}_k^H\vec{z}\vec{z}^H\bar{\vec{u}}_k}{\bar{\vec{u}}_k^H\bar{\vec{u}}_k}\\
		=&-\dfrac{\vec{z}^H\left(\hat{\vec{\Lambda}} - \bar{\lambda}_k\left(\sigma_{\text{s}}^2\right)\vec{I}_M\right)^{-1}\vec{z}\vec{z}^H\left(\hat{\vec{\Lambda}} - \bar{\lambda}_k\left(\sigma_{\text{s}}^2\right)\vec{I}_M\right)^{-1}\vec{z}}{\vec{z}^H\left(\hat{\vec{\Lambda}} - \bar{\lambda}_k\left(\sigma_{\text{s}}^2\right)\vec{I}_M\right)^{-2}\vec{z}}\\
		=&-\dfrac{1}{\sigma_{\text{s}}^4\vec{z}^H\left(\hat{\vec{\Lambda}} - \bar{\lambda}_k\left(\sigma_{\text{s}}^2\right)\vec{I}_M\right)^{-2}\vec{z}}\\
		=&-\dfrac{1}{\sigma_{\text{s}}^4\vec{a}^H\left(\hat{\vec{R}} - \bar{\lambda}_k\left(\sigma_{\text{s}}^2\right)\vec{I}_M\right)^{-2}\vec{a}}.\raisetag{4.5\baselineskip}
	\end{aligned}
\end{gather}
Taking the derivative of \eqref{eq:defObj} by applying the identity in \eqref{eq:derivStep} concludes our proof of \eqref{eq:derivUCF}.
\section{Deflation Process}\label{appsec:deflation}
	In this section, we describe the deflation process \cite[p. 471]{golub2013matrix}, \cite[Sec. 2]{Bunch1978} to simplify the eigenvalue decomposition in \eqref{eq:genericRankOneMod} where the initial diagonal matrix $\vec{D}=\text{diag}(d_1,\ldots,d_K)$ contains repeated eigenvalues, and there are several zero-valued entries in ${\vec{z} = \left[z_1, \ldots, z_K\right]^T}$.
	\begin{itemize}
		\item[(a)] If there exists an index $k$ such that $z_k = 0$, then the eigenvalue $\bar{d}_k$ in \eqref{eq:genericRankOneMod} is equal to $d_k$, since the $k$-th row and column of the diagonal matrix $\vec{D}$ are unperturbed by the rank-one matrix $\rho\vec{z}\vec{z}^H$. The remaining eigenvalues $\bar{d}_j$ with $j\neq k$ are the eigenvalues of $\hat{\vec{D}} - \rho\hat{\vec{z}}\hat{\vec{z}}^H$ where the diagonal matrix $\hat{\vec{D}}$ and the vector $\hat{z}$ are obtained by removing the $k$-th entry from the diagonal matrix $\vec{D}$ and the vector $\vec{z}$, respectively.
		\item[(b)] If there are two identical eigenvalues $d_k = d_i$ with $k\neq i$, we choose a Givens rotation matrix $\vec{G} = \left[\vec{g}_1, \ldots, \vec{g}_K\right]$ such that
		\begin{subequations}
		\label{eq:GivenChoice}
		\begin{align}
		\vec{g}_i^H\vec{z} &= \sqrt{\normsca{z_i}^2 + \normsca{z_k}^2}\\
		\label{eq:GivensNull}\vec{g}_k^H\vec{z} &= 0\\
		\vec{g}_j^H\vec{z} &= z_j \text{ with } j\neq i, j\neq k.
		\end{align}
		\end{subequations}
		Since $\vec{G}$ is unitary and $\vec{G}^H\vec{D}\vec{G} = \vec{D}$, the eigenvalues $\left\lbrace\bar{d}_1, \ldots, \bar{d}_K\right\rbrace$ of the original problem in \eqref{eq:genericRankOneMod} are identical to the eigenvalues of the matrix ${\vec{G}^H\left(\vec{D} - \rho\vec{z}\vec{z}^H\right)\vec{G} = \vec{D} - \rho\tilde{\vec{z}}\tilde{\vec{z}}^H}$ with $\tilde{\vec{z}} = \vec{G}^H\vec{z}$. However, the identity in \eqref{eq:GivensNull} implies $\tilde{z}_k = 0$, and therefore we can reduce this case to the case in (a).
	\end{itemize}
In the deflation process, the two above mentioned steps are applied iteratively to determine all the eigenvalues which remain unchanged due to the Hermitian rank-one modification, and to generate a deflated diagonal matrix $\hat{\vec{D}}$ with distinct eigenvalues and a deflated vector $\hat{\vec{z}}$ with non-zero entries. The remaining eigenvalues are then determined by applying Algorithm~\ref{alg:RootingSecularFunction} to the deflated matrix $\hat{\vec{D}} -\rho\hat{\vec{z}}\hat{\vec{z}}^H$.

%
%

\ifCLASSOPTIONcaptionsoff
  \newpage
\fi



%
\addcontentsline{toc}{chapter}{Bibliography}
\bibliography{references}{}
\bibliographystyle{ieeetran}

%

%
%
%




\end{document}